\documentstyle[preprint,aps,eqsecnum, epsfig]{revtex}
\begin{document}
\tighten
\preprint{LPTHE-99-32}
\draft 
\title{Non-equilibrium dynamics in quantum field theory at high
density: the tsunami}
\author{\bf F. J. Cao and  H. J. de Vega}
\address
{LPTHE, Universit\'e Pierre et Marie Curie (Paris VI) et Denis Diderot 
(Paris VII), Tour 16, 1er. \'etage, 4, Place Jussieu, 75252 Paris, Cedex 05, 
France}
\date{\today} 
\maketitle
\begin{center}
\end{center}
\begin{abstract} The dynamics of a dense relativistic quantum fluid out
of thermodynamic equilibrium is studied in the framework of the $
\Phi^4 $ scalar field theory in the large $ N $ limit.
The time evolution of a particle distribution in momentum space (the
tsunami) is computed. The effective mass felt by the particles in such
a high density medium equals the tree level mass plus the expectation
value of the squared field. The case of negative tree level squared
mass is particularly interesting. In such case dynamical symmetry
restoration as well as dynamical symmetry breaking can
happen. Furthermore, the symmetry may stay 
broken with vanishing asymptotic squared mass showing the presence
of out of equilibrium Goldstone bosons. We study these phenomena and
identify the set of initial   
conditions that lead to each case. We compute the equation
of state which turns to depend on the initial state. Although the
system does not thermalize, the equation of state for asymptotically broken 
symmetry is of radiation type. We compute the correlation
functions at equal times. The two point correlator for late times
is the sum of different terms. One stems from the initial particle
distribution. Another term accounts for the out of equilibrium
Goldstone bosons created by spinodal unstabilities when the symmetry
is asymptotically broken. Both terms are of the order of the inverse of
the coupling for distances where causal signals can connect the two points. The
contribution of the out of equilibrium Goldstones exhibits scaling
behaviour in a generalized sense.
\end{abstract}

\section{Introduction}

Physical phenomena in high energy density situations cannot be treated
with the usual perturbation methods. Self-consistent non-perturbative
methods are necessary in order to describe the out-of-equilibrium dynamics of
relaxation of quantum fields in such situations. The large $ N $ limit
is a particularly powerful tool for scalar models. The need for a
self-consistent method stems from the fact that the particle
propagation in such situations depends on the detailed state of the
system. 

Relevant physical situations at high energy densities arise in
the hadron (quark-gluon) plasma as in ultrarelativistic heavy ion
collisions, in the interior of dense stars and in the early
universe\cite{fen,infla}.  

We consider a dense relativistic quantum gas out of thermodynamic
equilibrium. We investigate the {\bf non-perturbative} dense regime in 
which the energy density  is of the order $ m_R^4/  \lambda $ where 
$ \lambda $ is the scalar self-coupling and $ m_R $ the physical mass in 
vacuum. In these conditions, even for very weakly coupled theories
non-linear effects are important and must be treated non-perturbatively.  
We thus use the large $ N $ limit that provides a non-perturbative 
scheme which respects the internal as well as the space-time symmetries,
is renormalizable and permits explicit calculations. More precisely,
we consider  the $ O(N) $ vector model with quartic self-interaction
and the scalar field in the vector representation of $ O(N) $. 

We study the physics of the in-medium effects at high energy
density. In such out of equilibrium regime the effective mass felt by
the particles is time-dependent and  different from the tree level
mass $ m_R^2 $. In the large-$N$ approximation, it takes the form
\begin{equation} \label{masefec}
{\cal M}^2(t) = m_R^2 + {\lambda \over 2 N}\langle {\vec \Phi}^2 \rangle (t)
\end{equation}
where $ {\vec \Phi}(x) $ stands for a $N$-component scalar field with
self-coupling  $ \lambda $.

We see that the quantum fluctuations of the scalar field 
directly contribute to the effective mass. We shall consider for
simplicity translationally invariant situations. That is, invariant
under  spatial traslations. Therefore, the effective mass and all
one-point expectation values depend on time but not on space coordinates. 

\bigskip

We work in the  small coupling
regime $ \lambda \ll 1 $. The reason being that the different
dynamical time scales are well separated in the $ \lambda \ll 1 $ regime. 
For larger couplings the dynamical time scales become of the 
same order and different physical phenomena get mixed up.

\bigskip

We choose as initial conditions a gaussian wave functional
and $ \langle \vec\Phi \rangle (0) = \langle \dot{\vec\Phi} \rangle(0) = 0 $.  
The calculation of correlation functions is done in the 
general case $ \langle {\vec\Phi} \rangle (0) \neq 0 $.

As initial distribution of particles we will choose a `tsunami' \cite{nos1}. 
That is, a distribution of particles in momentum
space that we choose for simplicity spherical. Typically, such
distribution is a shell peaked in a wavenumber 
$ k_0 $ with a large density of particles $ \sim m_R^3/\lambda $. 

We have seen that these conditions do not determine completely the initial 
state, and that the remaining degrees of freedom can be interpreted as 
the quantum coherence between different $k$-modes. We choose two highly 
coherent initial states that we will call case I and case II.

\bigskip

As we see in eq. (\ref{masefec}) the in-medium effects  give a 
{\bf positive} contribution to the effective mass. In general, this
contribution to the effective mass initially oscillates with time.
These oscillations can produce parametric 
resonance which is here shut-off after a few oscillations by the damping of 
the oscillations due to decoherence. This resonant effect is negligible
for small $ \lambda $. (Parametric unstabilities are shut off by the
backreaction for no particle initial conditions \cite{nos2,nos3,noscorre}). 

We present analytic solutions for initial particle distributions
narrow in momentum space. Such self-consistent solutions express in
close form in terms of elliptic functions, and reproduce the numerical
solutions with very good accuracy till the damping of the oscillations
becomes important (due to decoherence phenomena).
The narrower is the initial distribution in momentum, the longer in
time this effective solution holds.

For $ m_R^2 > 0 $, the effective mass is larger than the tree level mass
$ {\cal M}^2_\infty > m_R^2 $ (see table 1).

The more interesting case corresponds to $ m_R^2 < 0 $. We have in such case
spontaneously broken symmetry for low density and low energy states.
For large initial energy density, the positive definite term 
$ {\lambda \over 2}\langle {\vec \Phi}^2 \rangle (t) $ may overcome 
the negative squared mass $ m_R^2 < 0 $ in eq. (\ref{masefec}) and  the 
symmetry may be {\bf restored}: $ {\cal M}^2(t) > 0 $ 
(see table 2).

This happens at $ t = 0 $ for the case we call II provided the energy density
$ E $ fulfills 
$$ 
E > \frac{|m_R|^2}{\lambda_R}\,k_0^2 \quad .
$$
For late times the symmetry is restored in both cases, I and II, provided 
$$
E > \frac{|m_R|^4}{\lambda_R} + 2\,\frac{|m_R|^2}{\lambda_R}\,k_0^2
\quad .
$$
Moreover, we find in case II an interval of energies where the symmetry is 
initially unbroken ($ {\cal M}^2(0) > 0 $) and where it is broken
for asymptotic times: 
$$
 \frac{|m_R|^2}{\lambda_R}\,k_0^2 < E < 
\frac{|m_R|^4}{\lambda_R} + 2\,\frac{|m_R|^2}{\lambda_R}\,k_0^2 \quad .
$$
That is, we have a {\bf dynamical symmetry breaking} for these situations 
(see table 2).

\bigskip

We compute the asymptotic equation of state for this out of equilibrium 
system. 
That is, we derive for late times explicit formulae for the pressure
as a function of the total energy. The equation of state we obtain {\bf
explicitly} depends on the initial state, $ \lambda $ and  $ m_R $
(see table 3). Notice that even for 
asymptotic times the system does not thermalize. In particular, the
distribution functions reach non-thermal limits for $ t\to \infty $. 

When the symmetry is asymptotically broken, the equation of state takes
the radiation form 
$$
P = \frac{E}{3} \; .
$$
in spite of the fact that the system is out of equilibrium. Moreover,
the Goldstone theorem is valid here in this out of 
equilibrium situation. Namely, the effective squared mass $ {\cal M}^2(t) $  
asymptotically vanishes when the symmetry is asymptotically broken.

The equation of state for asymptotically unbroken symmetry 
[eqs. (\ref{pressurecaseI})] has the cold 
matter and the radiation equation of state as limiting cases.  

\bigskip

We explicitly compute the two point correlation functions.

For late times, the two point correlator at equal times $ C(|\vec{x}|,t) $ 
expresses as a sum of two or three terms:
\begin{equation}\label{correint}
C(|\vec{x}|,t) = C_{origin}(|\vec{x}|) + C_p(|\vec{x}|,t) + C_s(|\vec{x}|,t)
\end{equation}
There is the time-independent piece $ C_{origin}(|\vec{x}|) $ concentrated
around the origin. The pulse term, $ C_p(|\vec{x}|,t) $, is due to
the particles in the initial distribution that effectively propagate
as free particles with mass $ {\cal M}_\infty $.
$$
C_p(|\vec{x}|,t) =\frac1{\lambda |\vec{x}|} \; F(|\vec{x}|- 2\,v_g\,t-c) \; ,
$$
where $ v_g $ is the group velocity [$ v_g < 1 $ for unbroken
symmetry and $ v_g = 1 $ for dynamically broken symmetry where $ {\cal
M}_\infty  = 0 $]. $ F(u) $ is non-zero only around $ u = 0 $. 

The last term in eq. (\ref{correint}) is only present for dynamically broken
symmetry. 
$$
C_s(|\vec{x}|,t) = \frac{K}{\lambda \, |\vec{x}|} \; 
Q\!\left( \frac{|\vec{x}|}{2t} \right) \; .
$$
where $ K $ is a constant.
The function $ Q(u) $ is of the order $ {\cal O}(1) $ only for $ 0
< u < 1 $ due to causality. When the order parameter $ \langle
\vec\Phi \rangle (t) $ is identically zero we have $ Q(u) = \theta(1-u) $. 

When the order parameter is {\bf nonzero}, $ Q(u) $ oscillates with $ u $. At a
given time $ t $, the number of oscillations of $ Q(u) $ in
the interval  $ 0 < u < 1 $ 
equals the number of oscillations performed by the order parameter 
$ \langle \vec{\Phi} \rangle (t) $ from  time $ t = 0 $ till time $ t $. 
That is, scaling exists for $ \langle \vec{\Phi} \rangle (0) \neq 0 $ 
in a generalized sense since the function $ Q(u) $ changes each time  
$ \langle \vec{\Phi} \rangle (t) $ performs an oscillation. 
This is due to the appearance of an extra length scale,
the initial value of the order parameter.

\section{The model}

We consider  the $ O(N) $-invariant scalar field model with quartic
self-interaction\cite{nos1,FRW} with the scalar field in the vector  
representation of $ O(N) $. 

The action and Lagrangian density are given by,
\begin{eqnarray}
S  &=&  \int d^4x\; {\cal L}\; \; ,\label{action} \cr \cr
{\cal L}  &=&   \frac{1}{2}\left[\partial_{\mu}{\vec{\Phi}}(x)\right]^2
-\frac12\;m^2\;{\vec{\Phi}^2}
- \frac{\lambda}{8N}\left({\vec{\Phi}^2}\right)^2 \; . \label{potential} 
\end{eqnarray}

The canonical momentum conjugate to $ {\vec{\Phi}}(x)$ is,
\begin{equation}
\vec{\Pi}(x) = \dot{\vec{\Phi}}(x), 
\end{equation}
and the Hamiltonian is given by,
\begin{equation}\label{hamilt}
H = \int d^3x\left\{
\frac12 \vec{\Pi}^2(x)+\frac12\;[\nabla\vec{\Phi}(x)]^2+
\frac12\;m^2\; {\vec{\Phi}^2} +
\frac{\lambda}{8N}\left({\vec{\Phi}^2}\right)^2  \right\}.
\end{equation}

In the present case we will restrict ourselves to a translationally
invariant situation, {\em i.e.} eigenstates of  the total momentum operator. 
In this case  the order parameter $< {\vec \Phi}(\vec{x}, t) >$ is
independent of the spatial coordinates $ \vec{x} $ and only depends on time. 

The Heisenberg equations of motion for the field operator take the form
\begin{equation}\label{ecHei}
\left[ \partial^2 + m^2 + { \lambda \over 2 \, N } \vec{\Phi}^2(x)
\right] \vec{\Phi}(x) = 0
\end{equation}

The coupling $ \lambda $ is chosen to remain fixed in the large $ N $ limit.

It is  convenient to write the field in the Schr\"odinger picture as
\begin{eqnarray}
\vec{\Phi}(x)=(\sigma(x), \vec{\eta}(x) )  
= ( \sqrt{N} \phi(t)+\chi(x), \vec{\eta}(x) )
\label{split}
\end{eqnarray}
with $ \langle {\vec \eta}(\vec{x},t) \rangle =0 $
where $ \vec{\eta} $ represents the $ N-1 $ `pions', and $ \phi(t) =
\langle \sigma(x) \rangle $. Thus, $ \langle \chi(x) \rangle = 0 $.

\subsection{The wave functional (Schr\"odinger picture)}

We shall consider Gaussian wave functionals of the
type\cite{nos1,FRW,losala,losala87,nos2} 

\begin{equation}
\Psi[{\vec \Phi}(.,t)]={\cal{N}}^{1/2}(t)\; \Pi_{\vec{k}} \exp
\left[-\frac{A_{\vec k}(t)}{2}\;
\vec{\eta}_{\vec{k}}\cdot\vec{\eta}_{-\vec{k}}\right]\;.
\label{wavefunct}
\end{equation}
where
\begin{equation}
{\vec \eta}_{\vec{k}}(t)=  \int d^3 x \; {\vec \eta}(\vec{x},t)\;
e^{i\vec{k}\cdot\vec{x}} 
\label{fourier1}
\end{equation}
[Hence,  we can assume $ A_{-\vec k}(t) = A_{\vec k}(t) $ without loss
of generality]. 

As shown below, such Gaussian wave functionals are solutions of the
Schr\"odinger equation in the $ N = \infty $ limit. 
The Hamiltonian (\ref{hamilt}) in the large $ N $ limit is
essentially a harmonic oscillator Hamiltonian with self-consistent,
time-dependent frequencies:

\begin{eqnarray} \label{hamilt2}
H(t) &=& N\, V\,h_{cl}(t) -\frac{\lambda}{8 \, N} \left( \sum_{\vec k}
\langle \vec{\eta}_{\vec{k}}\cdot\vec{\eta}_{-\vec{k}} \rangle
\right)^2 \\ &+&\sum_{\vec{k}}\left[
-\frac{1}{2}\frac{\delta^2}
{\delta\vec{\eta}_{\vec{k}}\cdot\delta\vec{\eta}_{-\vec{k}}}
+\frac{1}{2}\;\omega_{{\vec k}}^2(t)\; \vec{\eta}_{\vec{k}}\cdot
\vec{\eta}_{-\vec{k}}\right] \; , \nonumber
\end{eqnarray}
where $ h_{cl}(t) $ stands for the classical Hamiltonian
$$
h_{cl}(t) =\frac12 \left[ {\dot\phi}^2(t)+ m^2 \phi^2(t) +
\frac{\lambda}{4}\phi^4(t) \right]\; ,
$$
and
\begin{equation}\label{omegadef}
\omega_{{\vec k}}^2(t) = k^2 + m^2 + {\lambda \over 2} 
\left[ \phi^2(t) + \frac{1}{N}\langle \vec{\eta}^{\,2}(x) \rangle\right]\; .
\end{equation} 

The functional Schr\"odinger equation is then given by ($\hbar=1$)
\begin{equation}
i\frac{\partial\Psi}{\partial t}=H\Psi\; . \label{timedep}
\end{equation} 
More explicitly,
\begin{equation}
i\dot{\Psi}[{\vec \Phi}]=\left\{ N\, V\,h_{cl}(t)-\frac{\lambda}{8 \, N} \left
( \sum_{\vec k} \langle \vec{\eta}_{\vec{k}}\cdot
\vec{\eta}_{-\vec{k}} \rangle \right)^2 +
\sum_{\vec{k}}\left[-\frac{1}{2}\frac{\delta^2} 
{\delta\vec{\eta}_{\vec{k}} \cdot \delta\vec{\eta}_{-\vec{k}}}
+\frac{1}{2}\;\omega_{{\vec k}}^2(t)\; \vec{\eta}_{\vec{k}}\cdot
\vec{\eta}_{-\vec{k}}\right] \right\}\Psi[{\vec \Phi}] \; ,  \label{schrod}
\end{equation}
which then leads to a set of differential equations for $ A_{{\vec k}}(t)$.

The evolution equations for $ A_{{\vec k}}(t) $ and $ {\cal{N}}(t) $
are obtained by taking the functional derivatives and
comparing powers of $ \eta_{\vec k} $ on both sides. We obtain the following
evolution equations
\begin{eqnarray}
i\dot{A}_{{\vec k}}(t) & = & A_{{\vec k}}^2(t)-\omega_{{\vec
k}}^2(t)\; ,\label{riccati}\\
{\cal N}(t) & = & {\cal N}(0) \exp\left\{-i\int^t_0{dt'
\left[ 2 N V h_{cl}(t') - \frac{\lambda}{4 \, N}
\left( \sum_{\vec k} \langle \vec{\eta}_{\vec{k}}\cdot
\vec{\eta}_{-\vec{k}} \rangle(t')\right)^2 +
N \sum_{\vec k}A_{\vec k}(t')
\right]}
\right\}\; ,
\label{norma}
\end{eqnarray}
with $ A_{{\vec k}}(t)= A_{R{\vec k}}(t)+i\, A_{I{\vec k}}(t) $.

The time dependence of the normalization factor
$ {\cal N}(t) $ is completely determined by that of the $ A_{{\vec
k}}(t) $ as a consequence of unitary time evolution. 

Using the expression for the wave functional (\ref{wavefunct}) we have,
\begin{eqnarray}
\langle\vec{\eta}_{\vec{k}}\cdot\vec{\eta}_{-\vec{k}}\rangle
&&=\frac{<\Psi|\; \vec{\eta}_{\vec{k}}\cdot\vec{\eta}_{-\vec{k}}\; |\Psi>}
{<\Psi|\Psi>} \nonumber\\
&&=\frac{\int\int{ \; 
\prod_{\vec q} {\cal D}\vec{\eta}_{\vec{q}} \;
e^{-A_{R\vec q}(t) \; \vec{\eta}_{\vec q}\cdot\vec{\eta}_{-\vec q}}
\;  (\vec{\eta}_{\vec{k}}\cdot\vec{\eta}_{-\vec{k}}) }}
{\int\int{ \prod_{\vec q} {\cal D}\vec{\eta}_{\vec{q}} \;  
e^{-A_{R\vec q}(t)\;\vec{\eta}_{\vec q}\cdot\vec{\eta}_{-\vec q}}}} \nonumber\\
&&=\frac{\int{ {\cal D}\vec{\eta}_{\vec{k}}\;  {\cal D}\vec{\eta}_{\vec{-k}}
\; (\vec{\eta}_{\vec{k}}\cdot\vec{\eta}_{-\vec{k}})\; 
e^{-2A_{R\vec k}(t) \; \vec{\eta}_{\vec k}\cdot\vec{\eta}_{-\vec k}}}}
{\int{ {\cal D}\vec{\eta}_{\vec{k}}\;  {\cal D}\vec{\eta}_{\vec{-k}}\;  
e^{-2A_{R\vec k}(t)\;\vec{\eta}_{\vec k}\cdot\vec{\eta}_{-\vec k}}}}\nonumber\\
&&=\frac{N}{2\; A_{R{\vec k}}(t)}\;,
\end{eqnarray}
and
\begin{equation}
\langle\vec{\eta}_{\vec{k}}\cdot\vec{\eta}_{\vec{k'}}\rangle =
\langle\vec{\eta}_{\vec{k}}\cdot\vec{\eta}_{-\vec{k}}\rangle \; 
\delta_{\vec{k},-\vec{k'}} \quad ;
\end{equation}
leading to the self-consistency condition,
\begin{equation}
 \frac{\langle \vec{\eta}^{\,2}(x) \rangle}{N}  = \frac{1}{NV}\sum_{\vec
k}\langle\vec{\eta}_{\vec{k}}\cdot\vec{\eta}_{-\vec{k}}\rangle 
= \frac{1}{V} \sum_{\vec k} \frac{1}{2 A_{R{\vec k}}(t)}\;,
\end{equation}
or taking the infinite volume limit,
\begin{equation}
\frac{\langle \vec{\eta}^{\,2}(x) \rangle}{N}
={1 \over N} \int \frac{d^3 k}{(2\pi)^3}\; 
\langle\vec{\eta}_{\vec{k}}(t)\cdot \vec{\eta}_{-\vec{k}}(t)\rangle
=\int \frac{d^3 k}{(2\pi)^3} \frac{1}{2A_{R{\vec k}}(t)}\;.
\label{fluctAk}
\end{equation}

The expectation value of the Heisenberg equation of motion (\ref{ecHei}) yields
the equation of motion for the order parameter $ \phi(t) $\cite{nos1}:
\begin{equation}\label{ecmpo}
{\ddot \phi}(t) +m^2 \,
\phi(t)+\frac{\lambda}{2}\left[\phi^2(t)+\frac{\langle \vec{\eta}^{\,2}(x)  
\rangle}{N} \right] \phi(t)=0
\end{equation}
where we used that $ \langle \vec\eta \rangle = 0 $ and 
$ \langle \vec\eta^{\,2} \, \vec\eta \rangle = 0 $.

Eqs.(\ref{riccati})-(\ref{norma}) together with eq. (\ref{ecmpo}) define
the time evolution of this quantum state in the infinite $ N $ limit.

\subsection{The field modes (Heisenberg picture)}

The Ricatti equation (\ref{riccati}) can be linearized by writing $
A_{\vec k}(t) $ in terms of the functions $ \varphi_{\vec k}(t) $ as 

\begin{equation}
A_{\vec k}(t)=-i\frac{\dot{\varphi}^*_{\vec k}(t)}{\varphi^*_{\vec k}(t)} \; ,
\label{phidef}
\end{equation}
leading to the mode equations\cite{nos1}
\begin{equation}
\ddot{\varphi}^*_{\vec k}+\omega_{\vec k}^2(t)\;\varphi^*_{\vec k}=0\;.
\label{phidiff}
\end{equation}

The relation (\ref{phidef}) defines the mode functions $ \varphi_{\vec
k} $ up
to an arbitrary  multiplicative constant that we choose such that  the
wronskian takes the value, 
\begin{equation}
{\cal W}_{\vec k} 
\equiv \varphi_{\vec k}\dot\varphi^*_{\vec k} - 
\dot\varphi_{\vec k}\varphi^*_{\vec k} 
= 2 i \; . \label{wronskian}
\end{equation}

The functions $ \varphi_{\vec k} $ have a very simple interpretation: they
obey the Heisenberg equations of motion for the
pion fields obtained from the Hamiltonian (\ref{hamilt}). Therefore,
we can write the Heisenberg field operators as
\begin{equation}
\vec{\eta}(\vec x,t) = \int \frac{d^3{\vec k}}{\sqrt{2} (2\pi)^3}\;
\left[ \vec{a}_{\vec k}\; \varphi_{\vec k}(t) \;
e^{i\vec k \cdot \vec x}+\vec{a}^{\dagger}_{\vec k} \;\varphi^*_{\vec k}(t)\;
e^{-i\vec k \cdot \vec x} \right] 
\label{fieldexpansion}
\end{equation}
where $ \vec{a}_{\vec k} \; , \; \vec{a}^{\dagger}_{\vec k} $ are the 
time independent annihilation and creation operators with the usual 
canonical  commutation relations. 
Thus,  the $ \varphi_{\vec k}(t) $ are the mode functions of the field.

From eq. (\ref{phidef}) we obtain the following useful relations,
\begin{equation}
A_{R{\vec k}}(t) = \frac{1}{|\varphi_{\vec k}(t)|^2}\quad ,  \quad
\label{areal} A_{I{\vec k}}(t)  =  - \frac{1}{2} \frac{d}{d
t}\ln|\varphi_{\vec k}(t)|^2 \;.
\end{equation}
Then, using these relations and eq. (\ref{fluctAk}) we express 
the fluctuation in terms of the mode functions:
\begin{equation}
\langle\eta^2\rangle(t) \equiv \frac{\langle\vec{\eta}^{\,2}(x)\rangle}{N}
= \frac{1}{2} \int{ \frac{d^3 k}{(2\pi)^3}\; |\varphi_{\vec
k}(t)|^2 } \; . \label{fluctmode}
\end{equation}

\subsection{Definition of the particle number}

Since in a time dependent situation the definition of the particle number 
is not unique, we choose to {\em define} the particle number with respect to 
the eigenstates of the instantaneous Hamiltonian (\ref{hamilt2}) at
the {\em initial time}, {\em i.e.}
\begin{equation}
\hat{n}_{\vec k} = 
\frac{1}{N \omega_{\vec k} }\left[-\frac{1}{2}
\frac{\delta^2}{\delta\vec{\eta}_{\vec{k}}\cdot\delta\vec{\eta}_{-\vec{k}}}+
\frac12 \; \omega_{\vec k}^2  \; 
\vec{\eta}_{\vec{k}}\cdot\vec{\eta}_{-\vec{k}}\right]- \frac{1}{2}
\label{opnumberofpart} 
\end{equation}

Here, $ \omega_{\vec k}  $ is the  frequency (\ref{omegadef})
evaluated at $ t=0 $. 

The expectation value of the number operator in the time evolved state is then 
\begin{equation}
n_{\vec k}(t)=\langle \Psi|\hat{n}_{\vec k}|\Psi \rangle
=\frac{[A_{R{\vec k}}(t)-\omega_{\vec k} ]^2+A_{I{\vec
k}}^2(t)}{4\,\omega_{\vec k} \,A_{R{\vec k}}(t)} \\ 
=\frac{\Delta_{\vec k}^2(t)+\delta_{\vec k}^2(t)}{4[1+\Delta_{\vec
k}(t)]}\; ,\label{numbdk} 
\end{equation}

where $ \Delta_{\vec k}(t) $ and $\delta_{\vec k}(t) $ are defined through the
relations,  
\begin{equation}
A_{R{\vec k}}(t)=\omega_{\vec k} \;[1+\Delta_{\vec
k}(t)]\quad , \quad A_{I{\vec k}}(t)=\omega_{\vec k}  
\;\delta_{\vec k}(t)\;. \label{deltadef}
\end{equation}

In terms of the mode functions $ \varphi_{\vec k}(t) $ and
$ \dot{\varphi}_{\vec k}(t) $, the expectation value of the number operator
is given by  
\begin{equation}
n_{\vec k}(t)=\frac{1}{4 \; \omega_{\vec k} }\;
\left[|\dot{\varphi}_{\vec k}(t)|^2
+\omega^2_{\vec k} \,|\varphi_{\vec
k}(t)|^2\right]-\frac{1}{2}\;.\label{numbmode} 
\end{equation}
For initially broken symmetry the frequencies in
eq. (\ref{opnumberofpart}) are modified for low $ k $ according to
eq. (\ref{unsfrequ}). 

\subsection{Initial conditions}

We will take initial conditions for the field expectation value of the form,
\begin{equation}
\phi(0)=\phi_0 \;\; , \;\; \dot\phi(0)=0 \; ,
\end{equation}
since we can make $ \dot\phi(0) $ to vanish by a shift in time.

The initial quantum state defined by eq. (\ref{wavefunct}) is
determined giving $ A_{R{\vec k}}(0) $ and
$ A_{I{\vec k}}(0) $  for all $ \vec k $ or equivalently, using the relations
(\ref{deltadef}),  giving $ \Delta_{\vec k} \equiv  \Delta_{\vec k}(0) $
and $ \delta_{\vec k}\equiv  \delta_{\vec k}(0) $. 

\begin{equation}\label{Ainic}
A_{R{\vec k}}(0)=\omega_{\vec k} \;[1+\Delta_{\vec k} ]\quad , \quad 
 A_{I{\vec k}}(0)=\omega_{\vec k}  \;\delta_{\vec k} \;.
\end{equation}

We can invert eq. (\ref{numbdk}) expressing $ \Delta_{\vec k}(t) $ in
terms of $ \delta_{\vec k}(t) $ and the particle number $ n_{\vec k}(t) $,
\begin{equation}
\Delta_{\vec k}(t)=2\left[n_{\vec k}(t)\pm\sqrt{n_{\vec k}(t)^2+n_{\vec k}(t)-
\frac{\delta_{\vec k}(t)^2}{4}} \right]\;.\label{Deltak}
\end{equation}

Thus, we can express the initial conditions in terms of the initial
distribution of particles $ n_{\vec k}(0) $ and $ \delta_{\vec k}  $. 
However, there are {\bf two} possible values of $ \Delta_{\vec k}  $
for each $ n_{\vec k}(0) $ due to the $ \pm $ sign in eq. (\ref{Deltak}).
We shall call case I to the upper sign in eq. (\ref{Deltak}) and case
II to the lower sign. 
(It is very important to notice that this sign is {\bf not} the sign of 
$ m_R^2 $. But is due to the fact that the value of $ n_{\vec k} \neq 0 $ and 
$ \delta_{\vec k} $ do not fix completely the initial state for the modes.) 

The quantity $\delta_{\vec k}(t)$ appears as the phase of the wave
functional and will be chosen for simplicity to be zero at $t=0$,
\begin{equation}
\delta_{\vec k} =0 \; . \label{inidelta}
\end{equation}
For simplicity, we will consider spherically symmetric particle distributions
and with gaussian form:
\begin{equation}\label{inidist}
n_{\vec k}(0)=\frac{{\hat N}_0}{I}\exp\left[-
\frac{(k-k_0)^2}{\hat{\sigma}^2}\right]\; , 
\end{equation}
where  $ {\hat N}_0 $ is the total number of particles per unit volume
and $ I $ is a normalization factor. We shall always consider $ {\hat
N}_0 \gg m^3 $.   
As we shall see below (subsection \ref{caseI}) it will be convenient to 
consider $ {\hat N}_0 \sim m^3/g $.

As the initial conditions are spherically symmetric and the evolution equations
are invariant under rotations the solutions will be spherically symmetric. 
So, the dependence on $ \vec k $ is only through the  modulus $k$.

\bigskip

The initial conditions on the mode functions follow from the relation  
(\ref{phidef}) and  the initial conditions (\ref{Ainic}) on $
A_{\vec k}(t) $ plus the  wronskian constraint (\ref{wronskian}).
Thus, the initial conditions on the mode functions are:

\begin{equation}  \label{phini}
\varphi_k(0) = \frac{1}{\sqrt \Omega_k} \quad , \quad
\dot\varphi_k(0) = -\frac{i\, \Omega_k
+\omega_k \; \delta_k }{\sqrt\Omega_k} 
\; , \label{dphiini}
\end{equation}
with $\Omega_k$ defined by:
\begin{equation}
\Omega_k \equiv A_{Rk}(0)=\omega_k \;  [ 1 + \Delta_k 
]\;. \label{Omegadef} 
\end{equation}

\bigskip

One sees  that for $\delta_k =0$ and an initial state with no
particles ($ \Delta_k = 0 $), the initial conditions become: 
$\varphi_k(0)=1/\sqrt{\omega_k }$ and 
$\dot\varphi_k(0)=-i\sqrt{\omega_k }$.
These were the initial conditions used in refs. \cite{nos2,nos3,noscorre}.

\section{Evolution equations}

The evolution equations for the expectation value of the field $\phi(t)$ and
for the mode functions $\varphi_k(t)$, eq. (\ref{ecmpo}) and
 eq. (\ref{phidiff}) respectively, are:

\begin{eqnarray}
&&\frac{d^2\phi(t)}{dt^2}+\left\{m_B^2+\frac{\lambda}{2}\;[\phi^2(t)+
\langle\eta^2\rangle_B(t)]\right\} \phi(t)=0\; ,\label{zeromode}\\
&&\frac{d^2\varphi_k(t)}{dt^2}+\left\{k^2+m_B^2+\frac{\lambda}{2}\;[\phi^2(t)+
\langle\eta^2\rangle_B(t)]\right\}\varphi_k(t)=0\;,\label{kmodes}
\end{eqnarray}
with the self-consistent condition
\begin{equation}
\langle\eta^2\rangle_B(t) = \frac{1}{2} \int{ \frac{d^3k}{(2\pi)^3}\;
|\varphi_k(t)|^2 }\; .\label{selfcons}
\end{equation}

The above evolution equations must be renormalized. This is achieved by
demanding that  the equations of motion be finite. The divergent
pieces  are absorbed into a redefinition of the mass and coupling constant.
\begin{equation}
{\cal M}^2_d(t) \equiv
m_B^2+\frac{\lambda}{2}\;[\langle\eta^2\rangle_B(t)+\phi^2(t)]
=m_R^2+\frac{\lambda_R}{2}\;[\langle\eta^2\rangle_R(t)+\phi^2(t)] \; .
\label{renormass}
\end{equation}
A detailed derivation of the renormalization prescriptions requires a WKB
analysis of the mode functions $\varphi_k(t)$ that reveals their ultraviolet
properties. Such an analysis has been performed in refs.\cite{nos1,FRW,nos2}.  
In summary, quadratic and logarithmic divergences are absorbed in the
mass term while the coupling constant absorbs a logarithmically divergent piece
\cite{FRW,cosmo}. The renormalized quantum fluctuations take the form
\begin{equation}
\langle\eta^2\rangle_R(t) = \frac{1}{2} \int{ \frac{d^3k}{(2\pi)^3}\left[
|\varphi_k(t)|^2-\frac{1}{k}+\frac{\theta(k-\kappa)}{2k^3}\;
{\cal M}^2_d(t) \right]} \label{renofluc}
\end{equation}
with $\kappa$ an arbitrary renormalization scale.  

\bigskip

We define now dimensionless quantities choosing the physical mass in vacuum 
$ |m_R| $ as unit of mass:
\begin{eqnarray}  \label{dimless}
&&q \equiv \frac{k}{|m_R|} \; ;\;\;\;\; {\tau} \equiv |m_R|\;t \; ;\;\;\;\; 
\zeta^2(\tau) \equiv \frac{\lambda_R\; \phi^2(t)}{2|m_R|^2} \; ; \;\;\;\; 
\varphi_q(\tau) \equiv \sqrt{|m_R|} \; \varphi_k(t)\; ; \nonumber \\
&&g \equiv \frac{\lambda_R}{8\pi^2}\; ; \;\;\;\; 
\omega_q(\tau) \equiv \frac{\omega_k(t)}{|m_R|}\;
; \;\;\;\; \Omega_q \equiv \frac{\Omega_k}{|m_R|}\; ;   \nonumber\\
&& g\Sigma(\tau) \equiv \frac{\lambda_R}{2|m_R|^2}\,\langle \eta^2 \rangle_R(t)
\; .
\end{eqnarray}
In terms of these dimensionless quantities the equations of motion
(\ref{zeromode})-(\ref{kmodes}) become
\begin{eqnarray} 
&&\left[ \frac{d^2}{d\tau^2}\pm 1+\zeta^2(\tau)+g\Sigma(\tau) \right]  
\zeta(\tau)=0\;,                               \label{zeromodeeq} \\
&&\left[ \frac{d^2}{d\tau^2}+q^2\pm 1+\zeta^2(\tau)+g\Sigma(\tau) \right]
\varphi_q(\tau)=0 \; ;       \label{modeeq} \\
&&g\Sigma(\tau) =  g \int_0^\infty q^2dq \left\{
|\varphi_q(\tau)|^2 - \frac{1}{q} + \frac{\theta(q-1)}{2\, q^3 }
\,\frac{{\cal M}_d^2(\tau)}{|m_R|^2} \right\}\;,
\label{gsigma} 
\end{eqnarray} 
where we have chosen the renormalization scale $ \kappa = |m_R| $ for
simplicity and the sign $ \pm $ corresponds here to the sign of $ m_R^2 $. 

In eqs. (\ref{zeromodeeq})-(\ref{modeeq}),
\begin{equation} \label{mass}
{\cal M}^2(\tau) \equiv \frac{{\cal M}^2_d(\tau)}{|m_R|^2} =
\alpha+\zeta^2(\tau)+g\Sigma(\tau)
\end{equation}
plays the role of a time dependent effective squared mass. 
$ \alpha \equiv sign(m_R^2) = \pm 1 $.

Depending on whether $ {\cal M}^2(0) $ is positive or negative the symmetry
will be initially unbroken or broken.

\bigskip

The initial conditions become in dimensionless variables for 
$ {\cal M}^2(0) > 0 $:
\begin{eqnarray}
\zeta(0) &=& \zeta_0 \quad , \quad \dot \zeta(0) = 0 \; , \label{inieta} \\ 
\nonumber \\
\varphi_q(0) &=& \frac{1}{\sqrt \Omega_q} \quad , \quad
\dot \varphi_q(0) = -\frac{i\, \Omega_q+\omega_q \,\delta_q }
{\sqrt\Omega_q}\;, \label{inicondq}
\end{eqnarray}
with:
\begin{eqnarray}
\Omega_q &\equiv& \omega_q \; [1+\Delta_q ] \quad , \quad
\omega_q  = \sqrt{q^2+{\cal M}^2(0)}\;, \cr \cr
\Delta_q  &=& 2\left[n_q(0)\pm\sqrt{n_q^2(0)+n_q(0)-
\frac{\delta_q^2 }{4}} \; \right]\;.\label{Deltaq}
\end{eqnarray}
We call case I to the upper sign and case II to the lower sign.
(Notice that this sign is {\bf not} the sign of 
$ m_R^2 $. We have to distinguish cases I and II due to the fact that
the value of $ n_q(0) \neq 0 $ and $ \delta_q $ do not fix
completely the modes.)  

\bigskip

From (\ref{inidelta}) and (\ref{inidist}) we have:
\begin{equation}\label{distriI}
\delta_q  = 0 \quad , \quad 
n_q(0) =  \frac{N_0}{I} \exp \left[ -\frac{(q-q_0)^2}{\sigma^2} \right] \; . 
\end{equation}
with $ N_0 = \frac{\hat{N}_0}{|m_R|^3} $ and
$ \sigma = \frac{\hat{\sigma}}{|m_R|} $. 

\medskip

Therefore,
\begin{equation}\label{sigmaR}
g\Sigma(\tau) =  g \int_0^\infty q^2dq \left\{ |\varphi_q(\tau)|^2 - \frac1{q}
+ \frac{\theta(q-1)}{2q^3}\, {\cal M}^2(\tau) \right\}\;.
\end{equation}

To perform the numerical evolution, we have introduced a momentum cutoff 
$\Lambda$. The quantum fluctuations become for finite cutoff,
\begin{equation}
g\Sigma(\tau) = \frac{1}{1-\frac{g}{2}\log\Lambda}
\left\{\int_0^\Lambda q^2dq |\varphi_q(\tau)|^2 - \frac{g}{2}\,\Lambda^2
+ \frac{g}{2}\log\Lambda\;\left[\alpha+\zeta^2(\tau)\right]\right\}
+ O\left({g\over\Lambda^2}\right)\;.
\label{gsigmawithcutoff}
\end{equation}
This is a positive quantity [up to $ O(g) $]. $ \alpha = sign(m_R^2) = \pm 1 $.

\bigskip

In the case where the symmetry is initially broken ($ {\cal M}^2(0) <
0 $) the only change in eqs. (\ref{inieta})-(\ref{gsigmawithcutoff}) is that the
initial frequencies $ \omega_q $ are modified for low $ q $ as
follows\cite{nos3}: 
\begin{equation}
\omega_q = \left\{
\begin{array}{ll}
\sqrt{q^2 + |{\cal M}^2(0)|} \;  &  {\rm for} \;\; q^2 <
1 + |{\cal M}^2(0)|  \label{unsfrequ} \\
\sqrt{q^2 + {\cal M}^2(0)} \;  &  {\rm for} \;\; q^2 > 
1 + |{\cal M}^2(0)|  \label{stafrequ} \\ 
\end{array} \right.
\end{equation}

\section{Early time dynamics}

We study now the time evolution of a narrow particle distribution
$ n_q(0) $ peaked at $ q = q_0 $. We show below that we can
approximate the dynamics for early times using a {\bf single} mode
 $ \varphi_{eff}(\tau) $. We solve the time evolution of  $
\varphi_{eff}(\tau) $ in close form in terms of
elliptic functions. Moreover, we compare these analytic results with the full
numerical solution of eqs. (\ref{zeromodeeq})-(\ref{gsigma}).

For $ m_R^2 < 0 $, we only consider initial particle peaks with $ q_0^2 > 2 $.
Thus, they are well outside possible spinodally resonant bands.

For simplicity, we consider $ \zeta_0=0 $ in this section and in
sections \ref{latetime}, \ref{energysec} and \ref{eqofstate}, 

\subsection{Case I}	\label{caseI}

In this case we have: 
$ \Delta_q =2\, \left[n_{q}(0)+\sqrt{n_{q}^2(0)+n_{q}(0)}\;\right] $.

We see from eq. (\ref{Deltaq}) that the modes with $ n_q(0) \ll 1 $
have $ \Omega_q \simeq \omega_q  $. Therefore, 
$ \varphi_q(0) $ and $ \dot\varphi_q(0) $ are of order $ g^0 $ [see
eq. (\ref{inicondq})]. Thus, their contribution to $ g\Sigma(\tau) $
will be of order $ g $ for early times. 

On the other hand, we have for modes with $ n_q(0) = O(1/g) $ (as $ g \ll 1 $
this implies $ n_q(0) \gg 1 $)
\begin{equation}
\Delta_q \simeq  4 \; n_q(0)\gg 1 \quad ,  \quad 
\Omega_q \simeq 4 \; \omega_q   \; n_q(0)\gg 1 \; .
\label{OmegacaseI}
\end{equation}
Therefore, eqs. (\ref{inicondq}) imply that these modes have 
$ |\varphi_q(0)| \ll 1 $. 

Thus,
\begin{eqnarray}
g\Sigma_{I}(0) &=& O(g) \;. \\	 \label{gsigmI0}
{\cal M}^2(0) &=& \alpha + O(g) = \pm 1 + O(g) \;. \label{MI0}
\end{eqnarray}
where $ \alpha = {\rm sign}(m_R^2) = \pm 1 $.

For these modes with $ n_q(0) = O(1/g) $, $ |\dot\varphi_q(0)| \gg 1 $. 
Therefore,  these modes will then grow and their contribution will dominate 
$ g\Sigma(\tau) $ for early times.  

We  always consider that $ n_q(0) = O(1/g) $  for some interval in $ q $. 
Hence, its contribution to  
$ g\Sigma(\tau) $ will be of order one and will have an important effect 
on the dynamics. In such conditions, the total number of particles $
N_0 $ per unit volume is also of the order $ O(1/g) $.  

For the initial conditions considered (\ref{distriI}),
these dominant modes are in a small interval centered at $ q = q_0 $,
and they are in phase at least for short times. More precisely, the
$q$-modes  will stay in phase for   $ (q-q_0)\, \tau \ll 2\pi $. 
The modes which are in phase  contribute coherently to
$ g\Sigma(\tau) $ (each one with a contribution proportional to its
occupation number). 

Hence, for small $ \tau $ a good approximation is to consider that 
all the particles are in a single mode with $ q = q_0 $.
This approximation will apply as long as the modes with large occupation
number stay in phase. 

In these conditions eq. (\ref{gsigmawithcutoff}) yields,
\begin{equation} \label{apgsigI}
g\Sigma(\tau) = g\;  q_0^2\;  |\varphi_{eff}(\tau)|^2 \; \Delta q \; + O(g)
+ O(g\,\sigma) \; ,
\end{equation}
where $ \Delta q \approx \sigma $ [see eq. (\ref{distriI})].

Thus,
\begin{equation} \label{efdifeqI}
\frac{d^2 \varphi_{eff}} {d\tau^2} + \omega_{q_{0}}^2  \;
\varphi_{eff}(\tau) + g \; \Delta q \;  q_0^2 \;
|\varphi_{eff}(\tau)|^2 \; \varphi_{eff}(\tau) = 0  \; ,
\end{equation}
with $ \omega_{q_{0}}  = \sqrt{q_0^2+\alpha}\; $ 
[recall $ \alpha = {\rm sign}(m_R^2) = \pm 1 $ and we choose $ q_0^2 >
2 $].

We also have,
\begin{equation} \label{inicondefI}
N_0 = 4\pi\, q_0^2 \, \Delta q \;  n_{eff} \quad \Rightarrow \quad
\Omega_{eff} = \frac{N_0\;\omega_{q_{0}} }{\pi\, q_0^2\, \Delta q} \; .
\end{equation} 

Solving the non-linear differential equation (\ref{efdifeqI}) with the
initial conditions (\ref{inicondq}) and (\ref{inicondefI}), we obtain
$ \varphi_{eff}(\tau) $. 
We find from eq. (\ref{apgsigI}),
\begin{equation}	\label{gsigmacaseI}
g\Sigma_{I}(\tau) = (q_0^2+\alpha)\, \left(1+\frac{1}{1-2k^2}\right) 
\left[ \frac{1}{1-k^2\;
\mbox{sn}^2\left(\sqrt{\frac{q_0^2+\alpha}{1-2k^2}}\;\,\tau,
\;k\right)}- 1 \right] \;,
\end{equation}
with sn$(z,k)$ the Jacobi sine function and
\begin{equation}\label{modulo}
k = \sqrt{\frac{1}{2}\left(1-\frac{1}
{1+\frac{g\Sigma_{I\,max}}{q_0^2+\alpha}} \right) } \quad .
\end{equation}
The function (\ref{gsigmacaseI}) is non-negative and  oscillates 
between zero and
\begin{equation}	\label{gsigmaImax}
g\Sigma_{I\,max} = (q_0^2+\alpha)\, 
\left[ \sqrt{1+\frac{2gN_0}{\pi\,(q_0^2+\alpha)^{3/2} }} - 1 \right] \; ,
\end{equation}
with period
\begin{equation}
T = 2\sqrt{\frac{1-2k^2}{q_0^2+\alpha}} \; K(k) \; ,
\end{equation}
where $ K(k) $ stands for the complete elliptic integral of the first
kind.[Notice that eq. (\ref{modulo}) implies $ 1-2k^2 > 0 $].

\bigskip

\begin{figure}[h]
\epsfig{file=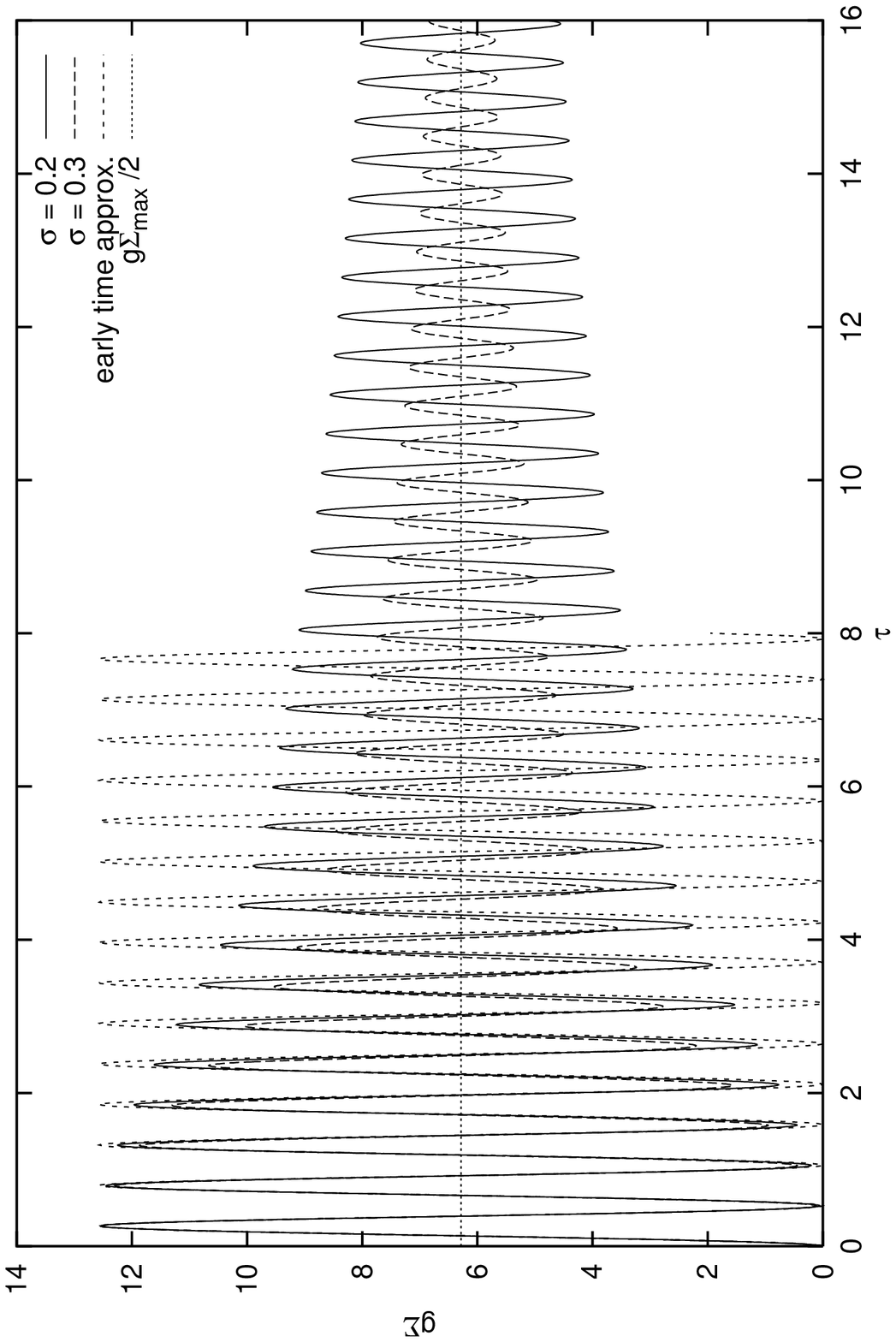}
\caption{Case I. $ m_R^2 > 0 $. Unbroken symmetry, $ gN_0 = 250 $, $ q_0 = 5 $,
$ \zeta_0 = 0 $, $ g = 10^{-7} $.
Comparison between numerical solutions and the early time approximation 
(\ref{gsigmacaseI}).}
\label{fcaseI}
\end{figure}

The elliptic solution (\ref{gsigmacaseI}) correctly
predicts  the amplitude of the first oscillation and the 
oscillation period (if the initial distribution of particles is not
too wide around $ q = q_0 $). See Fig. \ref{fcaseI}. 

The amplitude of the numerical solution is well reproduced by the
elliptic solution (\ref{gsigmacaseI}) till damping becomes
important. The oscillation period keeps well reproduced by
eq. (\ref{gsigmacaseI}) for longer times than the amplitude. [See
Fig. \ref{fcaseI}]. 
We see from the numerical solution of the exact equations
eqs. (\ref{zeromodeeq})-(\ref{gsigma}) that $ g\Sigma(\tau) $ exhibit
significatively damped oscillations after a few periods.  

The integral over  $ q $ for
$ g\Sigma(\tau) $ [eq. (\ref{sigmaR})] gets damped with time due to
the {\bf lost of coherence} between the different $q$-modes in the
distribution peak. We can apply here the adiabatic approximation (see
Appendix \ref{adiabatic}). In the late time limit, $ g\Sigma(\tau) $ therefore 
tends to the value $ g\Sigma_{I\,max}/2 $. In addition, the narrower is the
peak, the slower the oscillations in $ g\Sigma(\tau) $ are damped.
See Fig. \ref{fcaseI}.

The time scale where the numerical and the early times solution
deviate in amplitude,  essentially depends on the width $ \sigma $ of
the initial 
distribution. The smaller  is $ \sigma $, the latest the early time
solution (\ref{gsigmacaseI}) holds. Notice that 
eq. (\ref{gsigmacaseI}) gives $ g\Sigma(\tau) $ to zero order in $
\sigma $, whereas the damping of the oscillations is given by higher
orders in $ \sigma $.

In addition, we have seen from the numerical resolution that the
early time evolution depends on $ g $ only through $ gN_0 $ as predicted
by eqs. (\ref{modulo})-(\ref{gsigmaImax}).

\bigskip

When there is particle production through parametric resonance,
particles in the initial peak distribution are annihilated (in order to
conserve the total energy). This reduces the  contribution of the initial peak
to $ g\Sigma(\tau) $ whereas the oscillations due to the created
particles give a  contribution  to $ g\Sigma(\tau) $.
These oscillations are due to the coherence between the created 
particles at different $ q $.

Such changes in the distribution of particles influence the
asymptotic value of $ g\Sigma(\tau) $ for late times. Contrary to the
vacuum initial conditions, parametric resonance shuts off
here by the damping of the oscillations and not due to backreaction as
in ref.\cite{nos2}. Therefore, for small $ g $ the influence of
parametric resonances is highly suppressed. When parametric resonance 
is appreciable, it makes the dynamics to depend on $ g $ and {\em not only}
through the combination $gN_0$.

\bigskip

In case I, as $ {\cal M}^2(0) = \alpha = {\rm sign}(m_R^2) = \pm 1 $. 
The symmetry is initially spontaneously broken or not depending on 
whether $ m_R^2 $ (the squared tree level mass) is negative or positive.

\subsection{Case II}

In this case: $ \Delta_q  = 
2\, \left[n_{q}(0)-\sqrt{n_{q}^2(0)+n_{q}(0)}\;\right] $.

As in case I, $ g\Sigma(\tau) $ is dominated for short times by the
modes with large occupation numbers which are in phase at small $ \tau
$ due to the  initial conditions (\ref{distriI}).
Thus, we  can do the same approximation as in Case I, considering
that all particles are in a  single mode with $ q = q_0 $.

We have for the modes with $ q \simeq q_0 $, $ n_q(0) = O(1/g) \gg 1 $,
\begin{equation} \label{OmegacaseII}
\Delta_q \simeq -1 + \frac{1}{4 n_q(0)} \quad , \quad 
\Omega_q \simeq \frac{\omega_q }{4 n_q(0)} \ll 1 \; .
\end{equation}

Thus, for $ \tau = 0 $,
\begin{eqnarray}
g\Sigma_{II}(0)    &=& \frac{gN_0}{\pi\, \omega_{q_{0}} } \; 
+ O(g) + O(g\,\sigma) \;,		\label{gsigmaII0} \\ 
{\cal M}^2(0) &=& \alpha + g\Sigma_{II}(0) \;,\label{MII0}
\end{eqnarray}
where $ \alpha = {\rm sign}(m_R^2) = \pm 1 $.

Eq.(\ref{gsigmaII0}) is a third degree equation in $ g\Sigma_{II}(0)
$ since $ \omega_{q_{0}} = \sqrt{q_0^2 + \alpha + g\Sigma_{II}(0)} $.
The explicit solution is given in Appendix D. We find in the limiting
cases
\begin{eqnarray}\label{casolim}
g\Sigma_{II}(0)    &\buildrel{gN_0\gg
(q_0^2+\alpha)^{3/2}}\over=&\left(\frac{gN_0}{\pi}\right)^{2/3} 
-\frac13(q_0^2 +\alpha ) + {\cal
O}\left(\frac{q_0^2+\alpha}{gN_0}\right)^{2/3} 
\cr \cr
g\Sigma_{II}(0)    &\buildrel{gN_0\ll
(q_0^2+\alpha)^{3/2}}\over=& { gN_0 \over \pi \sqrt{q_0^2+\alpha}} + {\cal
O}\left({gN_0 \over q_0^2+\alpha}\right)^2
\end{eqnarray}

\medskip

For early times eq. (\ref{gsigmawithcutoff}) becomes
\begin{equation}
g\Sigma(\tau) = g\; q_0^2 \; \Delta q \; |\varphi_{eff}(\tau)|^2 \; 
+ O(g) +O(g\,\sigma) \; .
\end{equation}

Therefore,
\begin{equation}
\frac{d^2 \varphi_{eff}} {d\tau^2} + (q_0^2 + \alpha)\, \varphi_{eff}(\tau) 
+ g\; q_0^2\; \; \Delta q \; |\varphi_{eff}(\tau)|^2\;
\varphi_{eff}(\tau) = 0 \; , 
\label{diffeccaseII} 
\end{equation}

Using the initial conditions given by (\ref{inicondq}) and 
$ \delta_{eff}(0) = 0 $  we get from (\ref{OmegacaseII}),
\begin{equation}
\Omega_{eff} = \frac{\pi \; q_0^2 \; \Delta q \; \omega_{q_{0}} }
{N_0} \; , 
\end{equation}
where we have used that $ n_{eff} = \frac{N_0}{4\pi\, q_0^2\, \Delta q} $.

Thus, the solution of eq. (\ref{diffeccaseII}) with the specified initial 
conditions can be written as
\begin{eqnarray}	\label{gsigmacaseII}
g\Sigma_{II}(\tau) &=& g\Sigma_{I}(\tau+T/2) \cr\cr
&=& (q_0^2+\alpha)\, \left(1+\frac{1}{1-2k^2}\right) \left[ \frac{1}{1-k^2\; 
\mbox{sn}^2\left(\sqrt{\frac{q_0^2+\alpha}{1-2k^2}}\;\,(\tau+T/2),
\;k\right)}- 1 \right] \;,
\end{eqnarray}
where sn$(z,k)$ is the Jacobi sine function, $ T $ is the real period 
$$
T = 2\sqrt{\frac{1-2k^2}{q_0^2+\alpha}} \; K(k) \; ,
$$
$ K(k) $ stands for the complete elliptic integral of the first kind and
\begin{equation}
k = \sqrt{\frac{1}{2}\left(1-\frac{1}
{1+\frac{g\Sigma_{II\,max}}{q_0^2+\alpha}} \right) } \quad .
\end{equation}

The expression for $ g\Sigma_{II\,max} $ is here
\begin{equation} 	\label{gsigmaIImax}
g\Sigma_{II\,max} = g\Sigma_{II}(0) 
= \frac{gN_0}{\pi\,\sqrt{q_0^2+\alpha+g\Sigma_{II}(0)}}  \; .
\end{equation}

Notice that the relation $ g\Sigma_{II}(\tau) = g\Sigma_{I}(\tau+T/2) $
is true for given values of $ q_0 , \; \alpha $ and $ k $. 

Here $ g\Sigma(\tau) $ oscillates between zero and $ g\Sigma_{II\,max} $.
While the effective squared mass  $ {\cal M}^2(\tau) $ oscillates
between its initial value $ {\cal M}^2(0) = \pm 1 + g\Sigma_{II\,max} $ 
and its tree level value $ \pm 1 $ at the minima of $ g\Sigma(\tau) $.

\bigskip

We see that this approximation gives us correctly the amplitude of the 
first oscillation and the oscillation period; but not the damping of the
oscillation that is due to the dephasing  of the initial particle
distribution. The broader is the initial particle distribution, the more
effective the dephasing works and the faster the damping occurs.

As in case I, the smaller is $ \sigma $ 
(width of the initial particle distribution), the later the
early time solution (\ref{gsigmacaseII}) holds (see figures 2 and 3). Recall
that the damping of the oscillations is given by higher orders in $
\sigma $ while eq. (\ref{gsigmacaseII}) gives $ g\Sigma(\tau) $ to zero
order in $ \sigma $. These higher orders in $ \sigma $ will also break the
relation: $ g\Sigma_{II}(\tau) = g\Sigma_{I}(\tau+T/2) $; that we have 
find at zeroth order.

\bigskip

In case II, $ {\cal M}^2(0) = \alpha + g\Sigma_{II\,max} $. 
[Recall $ \alpha = {\rm sign}(m_R^2) = \pm 1 $.]
As $ g\Sigma_{II\,max} \ge 0 $, for $ m_R^2 > 0 $ the symmetry is initially
 unbroken. Instead, for $ m_R^2 < 0 $, the symmetry
is initially  unbroken for $ g\Sigma_{II\,max} > 1 $. 
For $ g\Sigma_{II\,max} < 1 $ the symmetry is initially spontaneously 
broken.

\section{Intermediate and late time dynamics}	\label{latetime}

We discuss here the intermediate and late time behavior of the quantum 
fluctuations and the effective squared mass. 
We mean by late time, times later than the spinodal time and than the 
damping time. We consider $ \zeta_0 = 0 $.

\bigskip

We observe from the numerical results the following common features in
the late time behavior for a wide range of initial conditions:

The asymptotic constant values of the magnitudes (mass, pressure,
number of particles) depend on $ g $,  for small $ g $, only through
the combination $ gN_0 $. (Except when parametric resonance is important).

Energy is conserved to one part in $ 10^7 $ confirming the precision
of our numerical calculations.

We have also seen that the mass tends to its limiting value  oscillating
with  an  amplitude that decays at least as $ \sim 1/\tau $. 
A similar  asymptotic decay has been found in this model for zero particle
initial conditions and $\zeta(0) \neq 0 $\cite{nos3}.

\bigskip

In  section \ref{eqofstate} we derive the asymptotic  equation of
state.

\subsection{ $ m^2_R > 0 $ ($ \alpha = +1 $)}	\label{mrpos}

We have for cases I and II that,
\begin{equation}
{\cal M}^2(\tau) = 1 + g\Sigma(\tau) \; ,
\end{equation}
although $ g\Sigma(\tau) $ has a different expression in each case.

As $ g\Sigma(\tau) \ge 0 $ the symmetry is {\bf unbroken} 
for all $ \tau $.

\bigskip

The adiabatic approximation holds (see Appendix \ref{adiabatic})
since parametric resonance is negligible in the weak coupling regime
considered here. Hence, the asymptotic value of  
$ g\Sigma(\tau) $ is $ \frac{g\Sigma_{max}}{2} $ and the asymptotic
squared mass goes to (see Fig. \ref{fcaseI})
\begin{equation}	\label{Minfmpos}
{\cal M}^2_\infty = 1 +\frac{g\Sigma_{max}}2  > 0\;.
\end{equation}

This result is in good agreement with the numerical calculations.

We find  that the mass tends to this value oscillating
with  an  amplitude that decays at least as $ \sim 1/\tau $. 

The initial peak of particles becomes lower and wider, and the total number
of particles slightly decreases compared to its initial value.

\bigskip

\begin{tabular}{c|c|c|} \cline{2-3}	\label{tabmpos}
$ $ & $ $ & $ $ 						\\
$ m_R^2>0 $  	&  Case I  	&  Case II  			\\ 
$ $ &$ $ &$ $ 							\\ \hline

\vline\hfill $ $ & $ $ & $ $ 					\\
\vline\hfill ${\cal M}^2(0)$ \hfill & $+1$&  $+1+g\Sigma_{II\,max}$ \\ 
\vline\hfill$ $ &$ $ &$ $ 					\\ \hline

\vline\hfill $ $ & $ $ &$ $ 					\\ 
\vline\hfill  initial \hfill &$ $ &$ $ 				\\
\vline\hfill symmetry \hfill	&  unbroken	&  unbroken	\\ 
\vline\hfill  $ $	&		&			\\ \hline

\vline\hfill $ $ &$ $ &$ $ 					\\
\vline\hfill ${\cal M}_\infty^2$ \hfill &  $+1+\frac{g\Sigma_{I\,max}}2$  & 
$+1+\frac{g\Sigma_{II\,max}}2$					\\
\vline\hfill $ $ &$ $ &$ $ 					\\ \hline 

\vline\hfill $ $ & $ $ &$ $ 					\\ 
\vline\hfill late time &$ $ &$ $ 				\\
\vline\hfill symmetry \hfill	& unbroken 	&  unbroken	\\ 
\vline\hfill $ $		&		&		\\ \hline
\end{tabular}

\bigskip

\begin{centerline}
{TABLE 1. Initial and late effective mass and symmetry for $ m_R^2 > 0 $.}
\end{centerline}

\subsection{ $ m^2_R < 0 $ ($ \alpha = -1 $)}

We have for both case I and case II that,
\begin{equation}
{\cal M}^2(\tau) = -1 + g\Sigma(\tau) \;.
\end{equation}
Recall that in case I the symmetry is initially broken since $ \Sigma_I(0) = 0
$. In case II, $ \Sigma_{II}(0) = \Sigma_{II\, max} $ and the symmetry
is initially broken (unbroken) for $ g\Sigma_{II\,max} < 1 $ ($
g\Sigma_{II\,max} > 1 $). 

\bigskip

\begin{figure}[h]
\epsfig{file=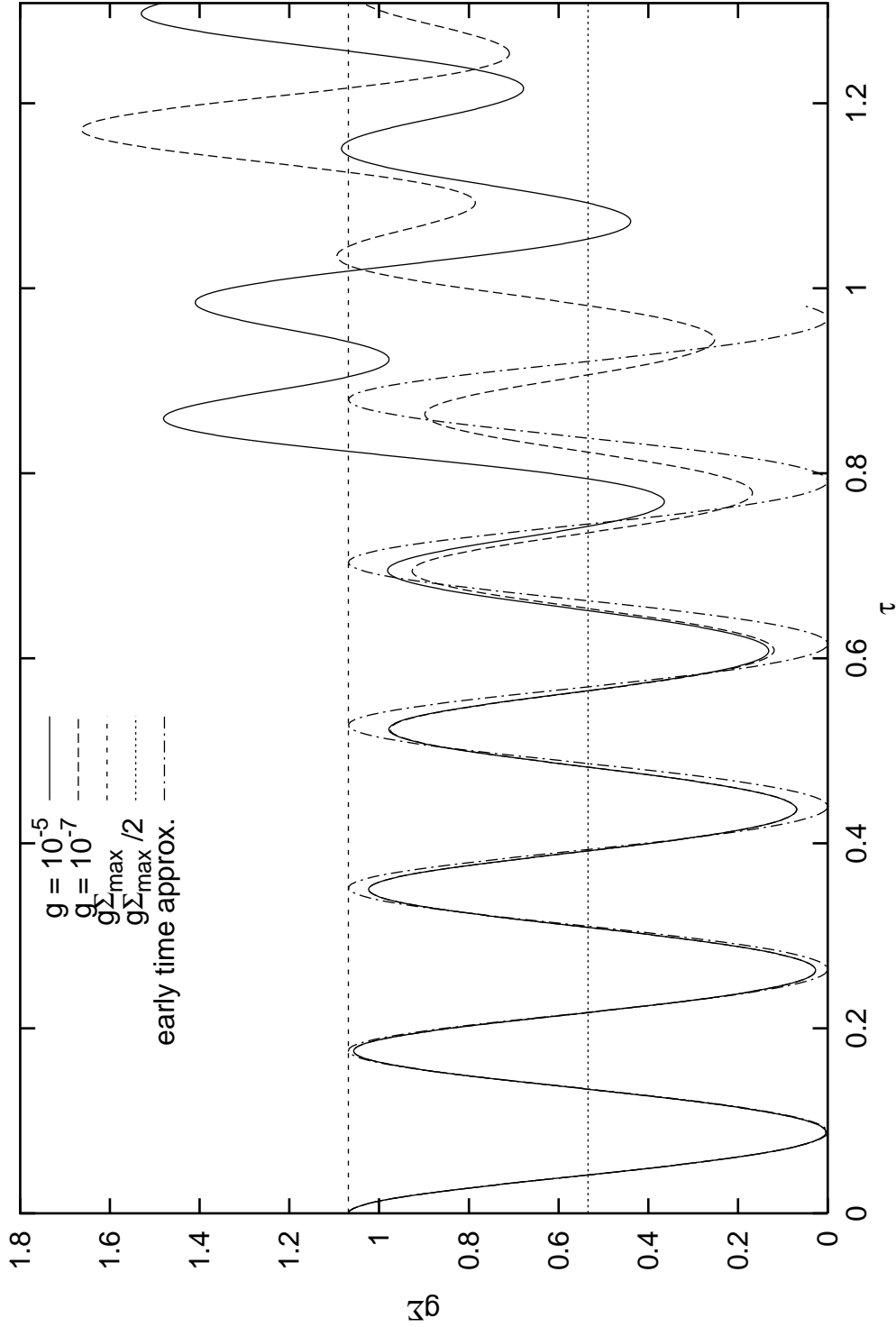}
\caption{Case II. $ m_R^2 < 0 $. $ g \Sigma(\tau) $ as a function of $ \tau $. 
Dynamically broken symmetry, $ gN_0 = 4.478 $, $ q_0 = 1.3083 $,
$ \sigma =  0.05233 $, $ \zeta_0 = 0 $.
Comparison between numerical solutions and the early time approximation 
(\ref{gsigmacaseII}). 
For late times $ g \Sigma(\tau) $ tends to $ 1 $, thus 
$ {\cal M}_\infty^2 = 0 $ [see Fig. (\ref{fcaseIIbinf})].  
\label{fcaseII}}
\end{figure}

\begin{figure}[h]
\epsfig{file=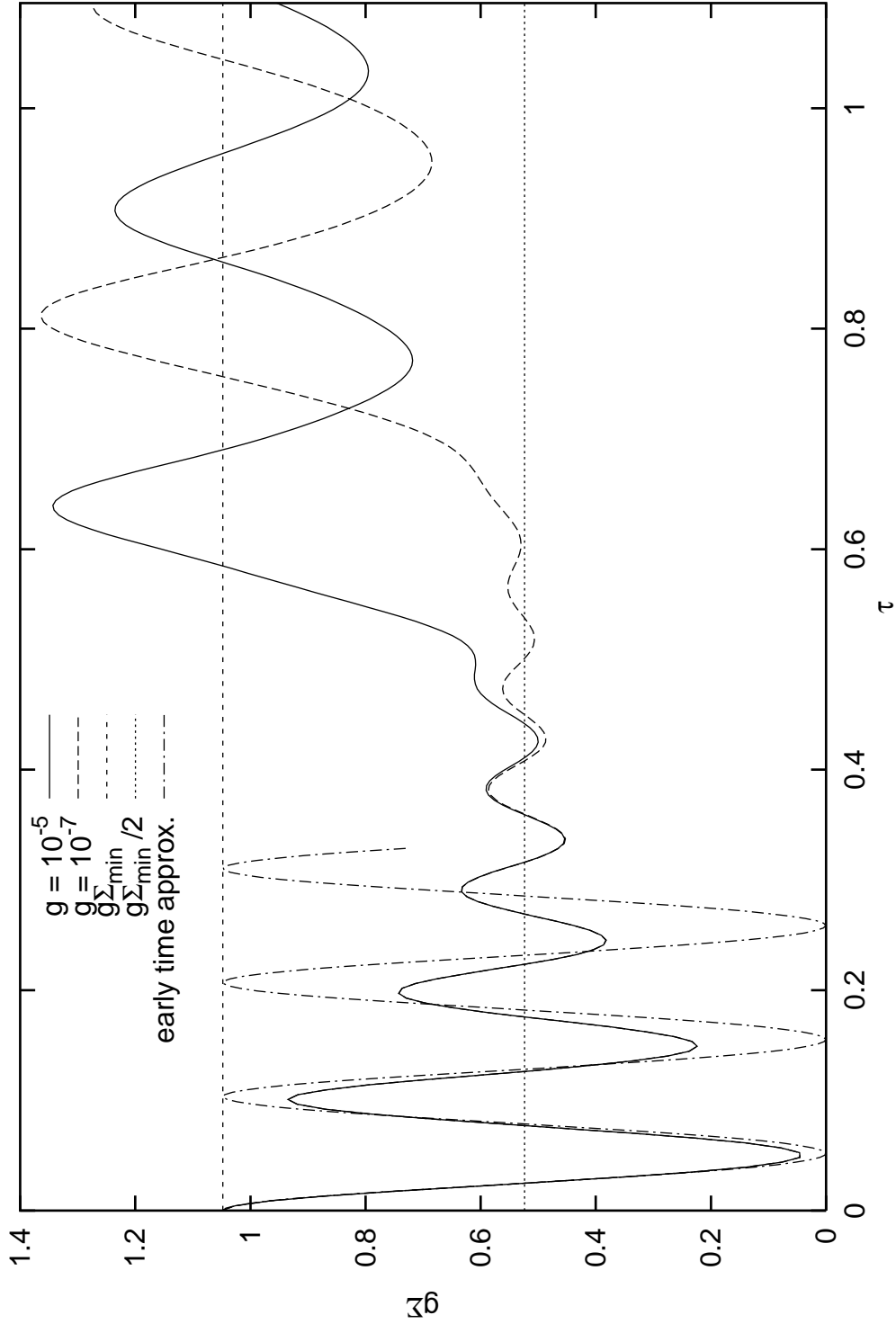}
\caption{Case II. $ m_R^2 < 0 $. $ g \Sigma(\tau) $ as a function of $ \tau $.
Dynamically broken symmetry, $ gN_0 = 5.101 $, $ q_0 = 1.5336 $, 
$ \sigma = 0.2191 $, $ \zeta_0 = 0 $.
Comparison between numerical solutions and the early time
approximation (\ref{gsigmacaseII}). For late times (not shown in the
figure) $ g \Sigma(\tau) $ tends to $ 1 $. \label{fcaseIIts}}
\end{figure}

\begin{figure}[h]
\epsfig{file=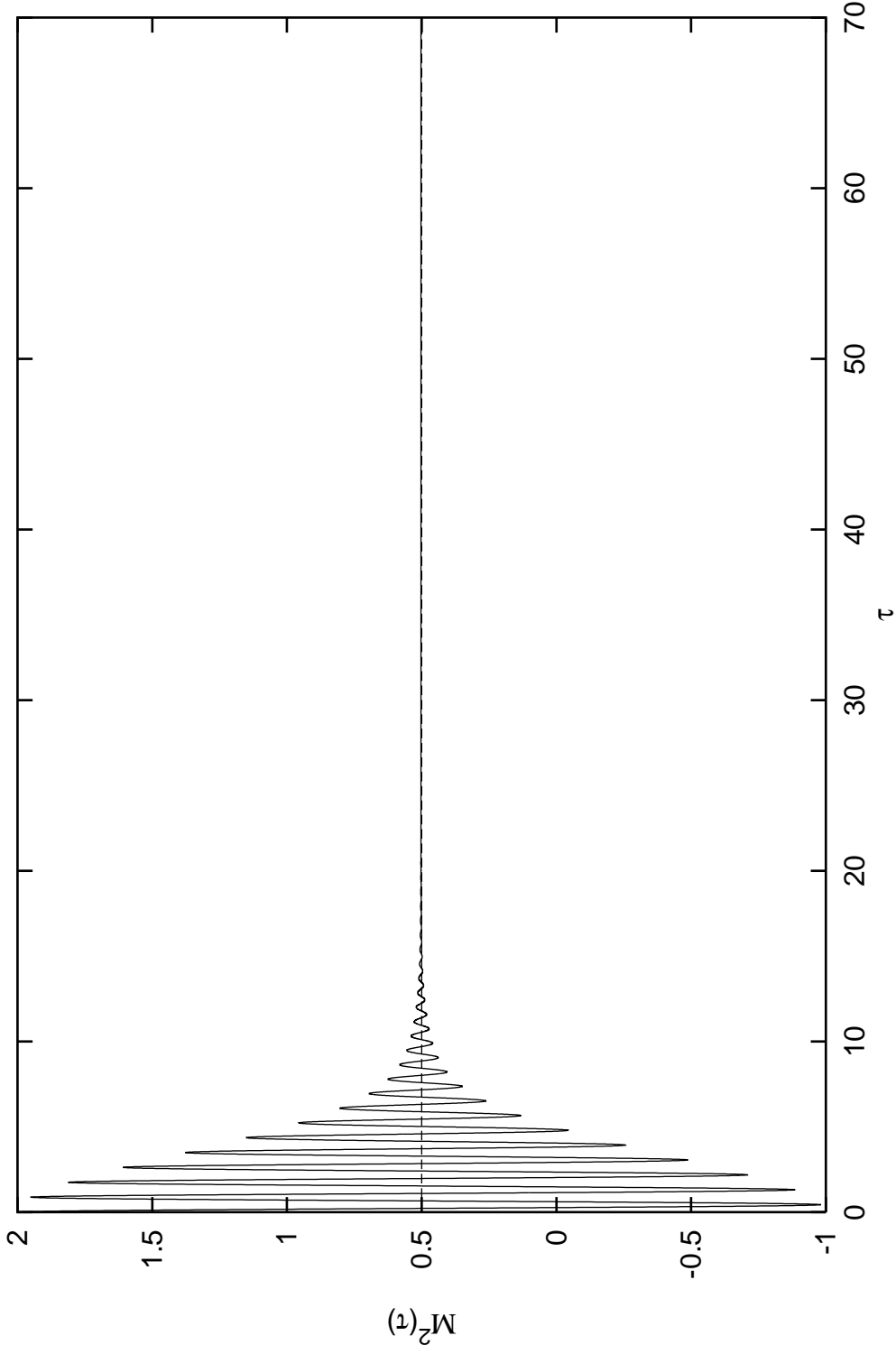}
\caption{Case II. $ m_R^2 < 0 $. $ {\cal M}^2(\tau) $ as a function of $\tau $.
The symmetry is unbroken. $ gN_0 = 67.96 $, $ q_0 = 7.071 $, 
$ \sigma = 0.4243 $, $ \zeta_0 = 0 $. Thus, $ g\Sigma_{II\,max} = 3.000 $
[see eq. (\ref{gsigmaIImax})]
and $ {\cal M}_\infty^2 = -1+\frac{g\Sigma_{II\,max}}{2} = 0.5000 $ 
[see eq. (\ref{Minfmnegunbr})].}
\label{fcaseIIainf}
\end{figure}

\begin{figure}[h]
\epsfig{file=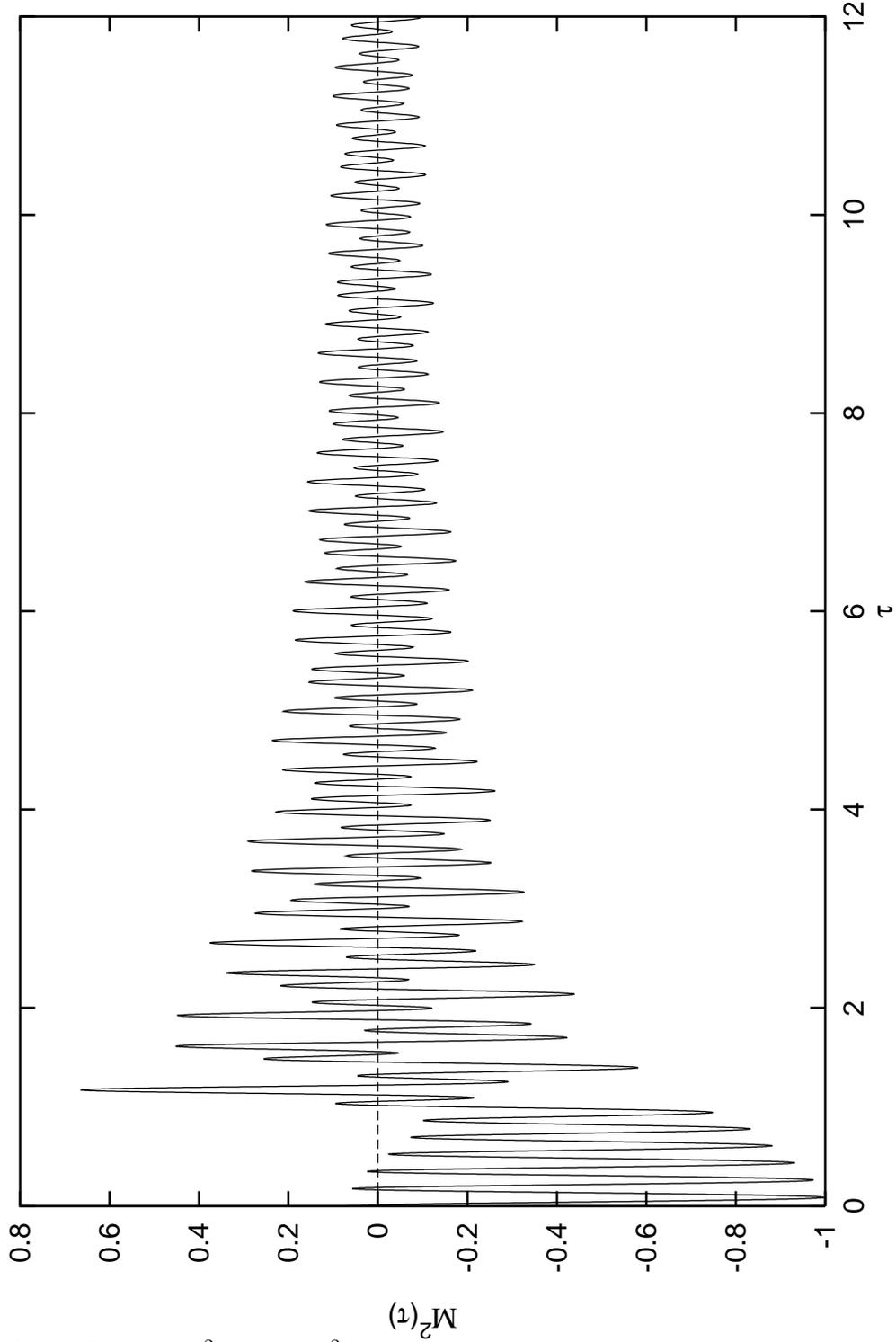}
\caption{Case II. $ m_R^2 < 0 $. $ {\cal M}^2(\tau) $ as a function of 
$ \tau $. The symmetry is dynamically broken. 
$ gN_0 = 4.478 $, $ q_0 = 1.3083 $, $ \sigma =  0.05233 $, $ \zeta_0 = 0 $.
(The same initial conditions and $ g $ as in Fig. \ref{fcaseII}.) }
\label{fcaseIIbinf}
\end{figure}

We have two different asymptotic regimes:

\begin{itemize}
\item $ \frac{g\Sigma_{max}}2 > 1 $: Asymptotically unbroken symmetry.

In this regime the results are similar to those in the previous subsection 
\ref{mrpos}, with
\begin{equation}	\label{Minfmnegunbr} 
{\cal M}^2_\infty = -1 +\frac{g\Sigma_{max}}2 > 0 \;.  
\end{equation}
This is indeed the value we obtain numerically (see Fig. \ref{fcaseIIainf}).

We find that for small $ g $ there is no appreciable spinodal
resonance here. This is so because $ {\cal M}^2(\tau) $ oscillates
around the positive value  $ {\cal M}^2_\infty $,
although for some intervals of time $ {\cal M}^2(\tau) < 0 $.

In this regime the  symmetry is {\bf restored} for late times due
to the presence of a high density of particles.

In particular, we have already noticed that for case I, $ {\cal M}^2(0) = -1$.
This does not change appreciably the dynamics and the symmetry gets restored
provided $ \frac{g\Sigma_{max}}2 > 1 $.

\item $ \frac{g\Sigma_{max}}2 < 1 $ Asymptotically broken symmetry.

In this regime for intermediate times (times earlier than the spinodal time
$ \tau_s $)  $ {\cal M}^2(\tau) $ oscillates around the negative value,
\begin{equation}	\label{mu2} \label{Minfmnegbr}
-\mu^2 = -1 +\frac{g\Sigma_{max}}2 < 0 \;.  
\end{equation}
giving rise to {\bf spinodal resonances}. (See figs. \ref{fcaseII}, 
\ref{fcaseIIts} and \ref{fcaseIIbinf}.) The dynamics turns to be
similar as for a constant squared mass  $ -\mu^2 $.
That is, the spinodally resonant band is in the $q$-interval from $ q
= 0 $ to $ q = \mu $ 
and the spinodal time $ \tau_s $ is the same as for a constant 
squared mass of $ -\mu^2 $ (see appendix \ref{secspinodaltime}).

\bigskip
We can further distinguish:
\begin{itemize}
\item when $ g\Sigma_{max} < 1 $, $ \; {\cal M}^2(\tau) $ is always negative 
for times $ \tau < \tau_s $.

\item when $ g\Sigma_{max} > 1$, $ \; {\cal M}^2(\tau) $ can be
temporarily positive due to oscillations. This happens in 
case I where $ \; {\cal M}^2(0) = -1 $.

In case II, $ \; {\cal M}^2(0) = -1 + g\Sigma_{max} > 0 $, but after a time of
order $ T/2\; , {\cal M}^2(\tau) < 0 $, and $ {\cal M}^2(\tau) $
continues to oscillate around the negative value given by eq. (\ref{mu2}).
Thus, {\bf in case II} we have  {\bf dynamical symmetry breaking}.

\end{itemize}

\bigskip

At time $ \tau_s $ the spinodal resonance has created enough particles 
(of the order $ O(1/g) $) to give an important contribution to $
g\Sigma(\tau) $.  
This finally makes $ g\Sigma(\tau) $ to oscillate around $ 1 $. 
Thus, $ {\cal M}^2(\tau) $ oscillates around zero and the spinodal resonance
stops. The particles created by the spinodal resonance are coherent. This
gives new oscillations to $ {\cal M}^2(\tau) $. These oscillations get damped
and the squared mass goes for late times to its asymptotic value,
\begin{equation}
{\cal M}_\infty^2 = 0 \;.
\end{equation}

The vanishing of the effective mass is accompanied by the presence of
Goldstone bosons out of equilibrium as in \cite{nos1,nos2}.

Particles created by spinodal resonances remain in the
$q$-interval from $ 0 $ to  $ \mu $.
This creation of particles depletes the initial peak of particles 
keeping  the total energy conserved.

For $ \tau < \tau_s $ the dynamic depends on $ g $, for small $ g $, only 
through the combination $ gN_0 $. However, $ \tau_s $ depends
explicitly on $ g $. 

\end{itemize}

\bigskip

\begin{tabular}{c|c|c|c|c|c|c|} \cline{2-7}	\label{tabmneg}
$  $  &  \multicolumn{2}{|c|}{} &  \multicolumn{4}{|c|}{} 	\\
$ m_R^2<0 $  	&  \multicolumn{2}{|c|}{Case I}	&  
\multicolumn{4}{|c|}{Case II} 					\\ 
$  $  & \multicolumn{2}{|c|}{} & \multicolumn{4}{|c|}{} 	\\ \hline

\vline\hfill $  $  & \multicolumn{2}{|c|}{} & \multicolumn{4}{|c|}{} \\
\vline\hfill ${\cal M}^2(0)$ \hfill &  \multicolumn{2}{|c|}{$-1$}	&  
\multicolumn{4}{|c|}{$-1+g\Sigma_{II\,max}$}			\\
\vline\hfill $  $  & \multicolumn{2}{|c|}{} & \multicolumn{4}{|c|}{} \\ \hline

\vline\hfill $ $ & \multicolumn{2}{|c|}{} & & \multicolumn{2}{|c}{} & \\
\vline\hfill initial \hfill	& \multicolumn{2}{|c|}{} 	&
if $g\Sigma_{II\,max}<1$ & \multicolumn{2}{|c}{$\;\;\;$}	&
if $g\Sigma_{II\,max}>1$ $\;\;\;$ 				\\
\vline\hfill symmetry 	&  \multicolumn{2}{|c|}{broken}	&  
broken			&  \multicolumn{2}{|c}{$\;\;\;$}	& 
unbroken $\;\;\;$ 						\\
\vline\hfill $ $ & \multicolumn{2}{|c|}{} & & \multicolumn{2}{|c}{} & \\ \hline

\vline\hfill $ $ & & & \multicolumn{2}{|c}{} & & 		\\
\vline\hfill $ $ &  if $\frac{g\Sigma_{I\,max}}2<1$  & 
if $\frac{g\Sigma_{I\,max}}2>1$ &  \multicolumn{2}{|c}{$\;\;\;$} & 
if $\frac{g\Sigma_{II\,max}}2<1$ $\;\;\;$ & if
$\frac{g\Sigma_{II\,max}}2>1$					\\
\vline\hfill ${\cal M}_\infty^2$ &  $0$	& $-1+\frac{g\Sigma_{I\,max}}2$ & 
\multicolumn{2}{|c}{$\;\;\;$}	& $0$ $\;\;\;$ 	& 
$-1+\frac{g\Sigma_{II\,max}}2$ 					\\ 
\vline\hfill $ $ & & & \multicolumn{2}{|c}{} & & 		\\ \hline 

\vline\hfill $ $ & & & \multicolumn{2}{|c}{} & & 		\\
\vline\hfill late time \hfill & & &\multicolumn{3}{|c|}{} &	\\ 
\vline\hfill symmetry	& broken 	& unbroken	&
\multicolumn{2}{|c}{$\;\;\;$}	& broken $\;\;\;$ 	& 
unbroken		 					\\ 
\vline\hfill $ $ & & & \multicolumn{2}{|c}{} & & 	\\\hline
\vline\hfill $ $ &      &  & & \multicolumn{2}{|c|}{} &         \\
\vline\hfill change of  &               & dynamical &
& \multicolumn{2}{|c|}{dynamical} &                             \\
\vline\hfill symmetry   & no change     & symmetry  & no change & 
\multicolumn{2}{|c|}{symmetry} & no change                      \\ 
\vline\hfill $ $        &               & restoration &         & 
\multicolumn{2}{|c|}{breaking} &                                \\ 
\vline\hfill $ $ &      &  & & \multicolumn{2}{|c|}{} &         \\ \hline
\end{tabular}

\bigskip

\begin{centerline}
 {TABLE 2. Initial and late effective mass and symmetry for $ m_R^2 < 0 $.}
\end{centerline}

\section{Energy}	\label{energysec}

After renormalization, taking dimensionless variables and introducing a 
momentum cutoff, the renormalized energy can be written as \cite{nos2}
\begin{eqnarray}	\label{energy}
E_{ren} &=& \frac{2|m_R|^4}{\lambda_R} \; \epsilon, \\
\epsilon &=& \frac{1}{2}\,\dot\zeta^2(\tau) + \frac{\alpha}{2}\,\zeta^2(\tau) 
+ \frac{\lambda}{4}\,\zeta^4(\tau) + \frac14\,\frac{1-\alpha}{2} \cr\cr
&+& \frac{g}{2} \int_0^{\Lambda}{q^2\; dq\left[|\dot\varphi_q(\tau)|^2
+ \omega_q^2(\tau)\,|\varphi_q(\tau)|^2\right]} 
- \frac14\,\left[g\Sigma(\tau)\right]^2 \; + O(g) \;,
\end{eqnarray}
where $ \alpha = {\rm sign}(m_R^2) $.

One can easily check that the energy is conserved using the
renormalized equations of motion (\ref{zeromodeeq})-(\ref{gsigma}).

\bigskip

Using the initial conditions (\ref{inieta}), (\ref{inicondq}),
(\ref{distriI}) and $ \zeta_0 = 0 $ we obtain for the energy at $ \tau=0 $ 
\begin{equation}
\epsilon = \frac14\,\frac{1-\alpha}{2} + \frac{g}{2} 
\int^{\Lambda}_0{q^2\; dq\left(\Omega_q + \frac{\omega_q^2}{\Omega_q}\right)}
- \frac14\,\left[g\Sigma(0)\right]^2 \; + O(g) \; . \label{ener0}
\end{equation}
This equation gives the energy in terms of the initial data.

$ \epsilon - \frac14\,\frac{1-\alpha}{2} $ is always positive because
we consider $ \zeta_0 = 0 $ and $ \dot\zeta_0 = 0 $. 

\subsection{Case I}

Let us consider an initial narrow distribution of particles which is 
peaked at $ q = q_0 $. Thus, we can use eqs. (\ref{OmegacaseI}) and
(\ref{ener0}) and consider all particles in a single mode with $ q =
q_0 $ as in previous sections. (Recall that in this case $
\omega_{q_0}^2 = q_0^2 + \alpha $.) 

The energy is then given by:
\begin{equation}	\label{energyI}
\epsilon = \frac{gN_0}{2\pi}\,(q_0^2+\alpha) 
+ \frac{1}{4}\,\frac{1-\alpha}{2} \; + O(g) + O(g\,\sigma) \;.
\end{equation}

\subsection{Case II} \label{caseII}

Under the same approximations [now using eq. (\ref{OmegacaseII})],
the energy is given by: 
\begin{equation}	\label{energyII}
\epsilon = (q_0^2 + \alpha)\,\frac{g\Sigma_{II}(0)}{2} 
+ \left[\frac{g\Sigma_{II}(0)}{2}\right]^2
+ \frac{1}{4}\,\frac{1-\alpha}{2} \; + O(g) + O(g\,\sigma)\;,
\end{equation}
where $ \omega_{q_0}^2 = q_0^2 + \alpha + g\Sigma(0) $ and
$ g\Sigma_{II}(0) $ is given by the relation (\ref{gsigmaII0}), {\it i.e.},
$ g\Sigma_{II}(0) = \frac{gN_0}{\pi\, \sqrt{q_0^2+\alpha+g\Sigma_{II}(0)}} $.

\subsection{$ g\Sigma_{max} $ in terms of the energy }

The energy and  $ g\Sigma_{max} $ have different expressions in cases
I and  II [see  eq. (\ref{gsigmaImax}) {\it vs.} eq. (\ref{gsigmaIImax})].  

However, it is important to notice that $ g\Sigma_{max} $ has the same
expression in terms of the energy {\bf both} for  cases I and II,
\begin{equation}	\label{gsigmamaxener}
g\Sigma_{max} = \sqrt{(q_0^2+\alpha)^2
+ 4\left(\epsilon-\frac14\,\frac{1-\alpha}{2}\right)} - (q_0^2+\alpha)
\; .
\end{equation}
We have verified  that eqs. (\ref{energyI})-(\ref{gsigmamaxener}) are
valid numerically for initial particle distributions with width $
\sigma < 1$. 

\section{Equation of state} 	\label{eqofstate}

We derive here the equation of state ({\em i.e.} the pressure as a function
of the energy) for asymptotic times. As shown below, the  asymptotic
equation of state {\bf depends} on the initial state.

We consider $ \zeta_0 = 0 $ and $ \delta_q = 0 $. 
Notice that $ \zeta_0 = 0 $ and $ \dot\zeta_0 = 0 $ implies 
$ \zeta(\tau) = 0 $ for all $ \tau $.

\bigskip

\subsection{Sum rule}

A further expression for the energy follows by  evaluating the right-hand side
of (\ref{energy}) in the $ \tau\to\infty $ limit. 
Using eqs. (\ref{approxsquaredmod}) and (\ref{approxsquaredmoddot}) yields
\begin{equation}	\label{enerinf}
\epsilon = \frac14\,\frac{1-\alpha}{2} +\int^\Lambda_0{q^4\;dq\;M^2_q(\infty)} 
+ {\cal M}_\infty^2\,g\Sigma_\infty - \frac14\,(g\Sigma_\infty)^2 \; + O(g)\;.
\end{equation}

[The cosine terms in eqs. (\ref{approxsquaredmod}) and 
(\ref{approxsquaredmoddot}) for late time are fastly oscillant in $ q $ 
and thus they do not contribute to the $ q$-integral in the infinite time
limit.]

Equation (\ref{enerinf}) allows us to express the integral over 
$ M^2_q(\infty) $ in terms of known quantities: the initial data and 
$ {\cal M}^2_\infty $,
\begin{equation}	\label{sumrule}
\int^\Lambda_0{q^4\;dq\; M^2_q(\infty)} =
\epsilon - \frac14\,\frac{1-\alpha}{2} - {\cal M}_\infty^2\,g\Sigma_\infty
+ \frac14\,(g\Sigma_\infty)^2 \; + O(g)\;.
\end{equation}
 
This sum rule holds for $ \zeta_0 = 0 $ and $ \delta_q = 0 $.

\subsection{Pressure and equation of state}

The renormalized pressure can be written as
\begin{eqnarray}
P_{ren}(\tau) &=& \frac{2|m_R|^4}{\lambda_R} \; p(\tau), \\
p(\tau) &=& -\epsilon +  \dot\zeta^2 + 
g\int^\Lambda_0{q^2\; dq\;\left[|\dot\varphi_q(\tau)|^2+
\frac{q^2}{3}|\varphi_q(\tau)|^2\right]} + O(g) \;.
\label{pressure}
\end{eqnarray}
We analogously evaluate the pressure in the infinite time limit with
the result,
\begin{equation}
p_\infty =  -\epsilon + \frac{4}{3} \int^\Lambda_0{q^4\;dq\;  M^2_q(\infty)}
+{\cal M}^2_\infty \, g\Sigma_\infty \;.
\end{equation}

Using now the sum rule (\ref{sumrule}) and $ {\cal M}_\infty^2 = \alpha 
+ g\Sigma_\infty $ yields 
\begin{equation}	\label{pinf}
p_\infty =  \frac{1}{3}\epsilon - \frac{\alpha}3\, g\Sigma_\infty
- \frac13\, \frac{1-\alpha}{2} \;.
\end{equation}
[Recall $ \alpha = {\rm sign}(m_R^2) $.]

\bigskip

We consider narrow distributions of particles centered at $ q = q_0 $.
To obtain the equation of state, we make the approximation of considering 
all particles in a single mode with $ q = q_0 $, and we express the pressure 
as a function of the energy.

The equation of state is {\bf the same for case I and case II}.

We have to distinguish between unbroken and broken symmetry for $ \tau
= \infty $:

\begin{itemize}
\item $ {\cal M}_\infty^2 > 0 $

In this regime $ g\Sigma_\infty=\frac{g\Sigma_{max}}{2} $ with 
$ g\Sigma_{max} $ given by eq. (\ref{gsigmamaxener}). 
Thus, the equation of state (\ref{pinf}) becomes
\begin{equation}	\label{eqofstunbroken}
p_\infty = \frac{1}{3}\, \epsilon - 
\frac{\alpha}{6} \left[ \sqrt{(q_0^2+\alpha)^2
+ 4\left(\epsilon-\frac14\,\frac{1-\alpha}{2}\right)} - (q_0^2+\alpha) \right] 
- \frac13\,\frac{1-\alpha}{2} \;. \label{pressurecaseI}
\end{equation}
This equation gives $ p_\infty $ as a function of $ \epsilon $ for an
initial momentum distribution centered at $ q_0 $. Therefore, the equation of
state explicitly depends on the initial conditions.

\medskip

Let us consider some limiting cases.

\begin{itemize}
\item  
In the limit, 
$ \epsilon - \frac14\,\frac{1-\alpha}{2} \ll (q_0^2+\alpha)^2 $, 
we have the equation of state,
\begin{equation}
p_\infty = \frac13\,\left(\epsilon - \frac14\,\frac{1-\alpha}{2}\right)
\left( 1 - \frac{\alpha}{q_0^2+\alpha} \right) - \frac14\,\frac{1-\alpha}{2}\;.
\end{equation}

For $ \alpha = +1 $ ($ m_R^2 > 0 $) this equation reduces to
\begin{equation}
p_\infty = \frac{\epsilon}{3} \left( 1 - \frac{1}{q_0^2 + 1} \right) \;,
\end{equation}
and interpolates between a cold matter (for $q_0\ll1$), and a
radiation (for $q_0\gg1$) equation of state.

\item 
In the opposite limit 
$ \epsilon - \frac14\,\frac{1-\alpha}{2} \gg (q_0^2+\alpha)^2 > 1 $
we have a radiation type equation of state $ p_\infty = \frac13\,\epsilon $.

\end{itemize}

\item $ {\cal M}_\infty^2 = 0 $

This can only happen if $ \alpha = -1 $ ($ m_R^2 < 0 $).  
$ {\cal M}_\infty^2 = -1 + g\Sigma_\infty = 0 $ implies
$ g\Sigma_\infty = 1 $ and we have a radiation type equation of state,
\begin{equation}	\label{eqofstbroken}
p_\infty = \frac{\epsilon}{3} \; .
\end{equation}

\end{itemize}

\bigskip

The pressure is continuous at the boundary between broken and unbroken 
symmetry, but its derivative with respect to the energy have a discontinuity 
$ \frac13\,\frac{1}{q_0^2+1} $. [See eq. (\ref{eqofstbroken}) {\it vs.} 
eq. (\ref{eqofstunbroken}).] 

\bigskip

We see that the equation of state is determined by the energy, the momentum 
$ q_0 $ around which the initial peak of particles is centered and the sign 
of the physical mass in vacuum $ m_R^2 $. But it does {\bf not} depend whether 
we are in case I or II. 

\medskip

\begin{tabular}{c|c|c|c|} \cline{2-4}		\label{tabeqofst}
 & & \multicolumn{2}{|c|}{}					\\
	& ${\cal M}_\infty^2=0$	& 
\multicolumn{2}{|c|}{${\cal M}_\infty^2>0$}			\\ 
 & & \multicolumn{2}{|c|}{}					\\ \cline{3-4}

	& 	& 		& 	 			\\ 
	& 	& $m_R^2<0$	& $m_R^2>0$ 			\\
	& 	& 		& 	 			\\ \hline

\vline\hfill $ $  & & &						\\ 
\vline\hfill equation & & &					\\ 
\vline\hfill of state \hfill & $p_\infty=\frac{\epsilon}{3}$ 	&
$ \; p_\infty=\frac{\epsilon}{3}+\frac{q_0^2-1}6\,\left[\sqrt{1+
\frac{4(\epsilon-\frac14)}{(q_0^2-1)^2}} -1\right] -\frac13 \; $ & 
$ \; p_\infty=\frac{\epsilon}{3}+\frac{q_0^2+1}6\,\left[\sqrt{1+
\frac{4\epsilon}{(q_0^2+1)^2}} -1\right] \; $			\\ 
\vline\hfill $ $ & & &						\\ \hline
\end{tabular}

\bigskip

\begin{centerline}
 {TABLE 3. Asymptotic equation of state in the different situations}
\end{centerline}

\section{Correlation functions and Bose condensate}

The equal time correlation function is given for an arbitrary time $
\tau $ by

\begin{equation}
\langle \eta^a(\vec{r},\tau) \eta^b(\vec{0},\tau) \rangle 
- \langle \eta^a(\vec{r},\tau) \rangle \; \langle \eta^b(\vec{0},\tau) \rangle
= \delta^{a,b} \;  C(\vec{r},\tau)
= \delta^{a,b} \int{\frac{d^3q}{2(2\pi)^3}\;  |\varphi_q(\tau)|^2 \;
e^{i\vec{q}\cdot\vec{r}}} \; .
\end{equation} 

The initial conditions  are specified in
eqs. (\ref{inieta})-(\ref{distriI}); and as they are rotationally invariant,  
\begin{equation}
C(r,\tau) = \frac{1}{4\pi
r}\int_0^{\infty}{q\;dq\;|\varphi_q(\tau)|^2\;\sin(qr)}\;\;.  
\end{equation}

In this section we consider $ \zeta_0 = 0 $ and $ \zeta_0 \neq 1 $
with $ \zeta_0 \ll 1 $.

\subsection{Early time}

We have the same results for $ \zeta_0 = 0 $ and for $ \zeta_0 \ll 1 $.

\subsubsection{Case I}

\begin{figure}[h] 
\epsfig{file=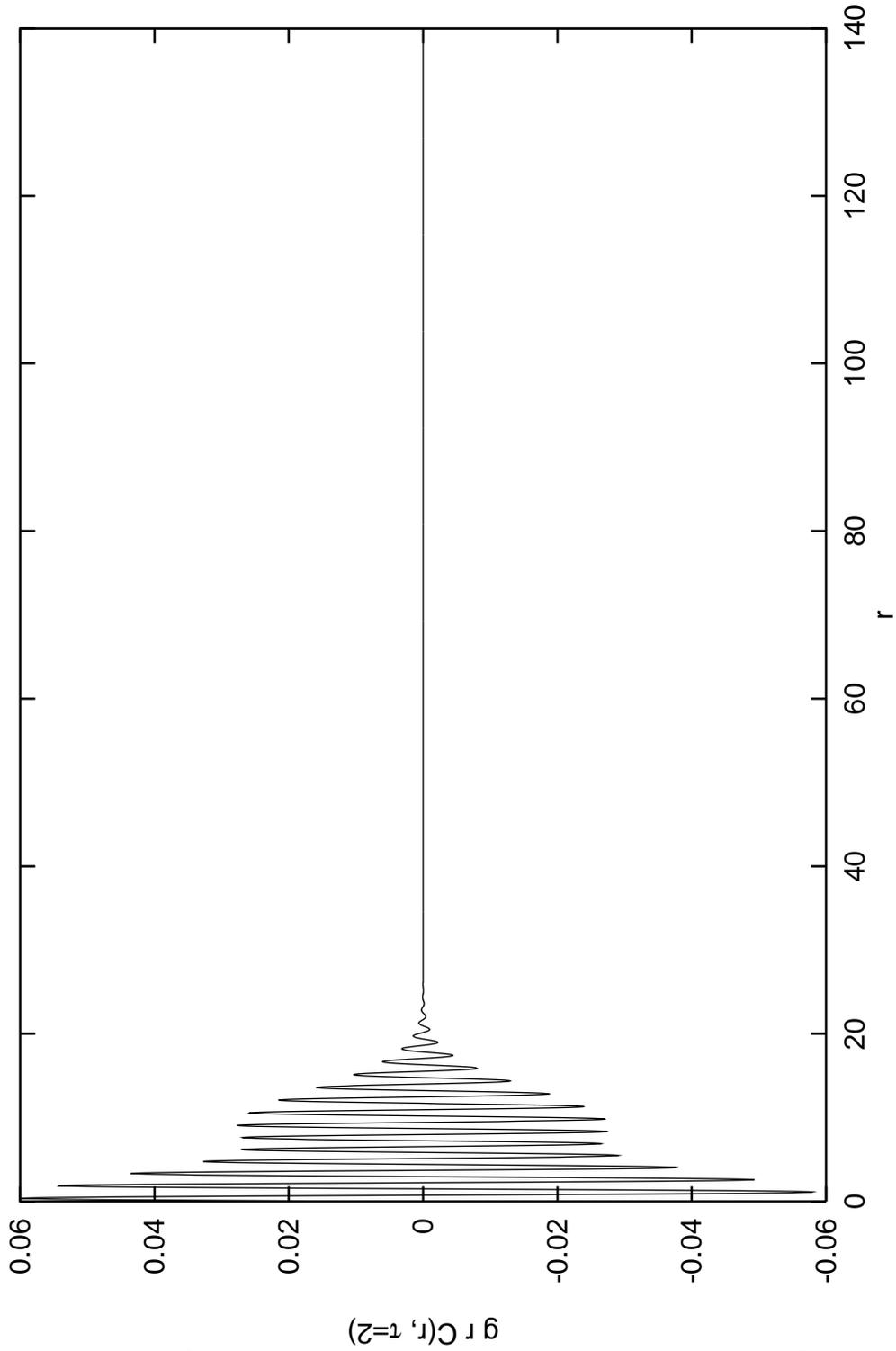}
\caption{Case I. $ m_R^2 > 0 $. Unbroken symmetry.
$ g r C(r, \tau=2) $ for $ g = 10^{-7} $ and initial conditions: 
$ \zeta_0 = 0 $, $ gN_0 = 250 $, $ q_0 = 4.0 $, $ \sigma = 0.3 $.}
\label{frcorrI2}
\end{figure}

At $ \tau = 0 $ the correlation functions is of order one, because 
$ |\varphi_q(0)|^2 \lesssim 1 $. [ $ |\varphi_q(0)|^2 \sim 1 $ for the
non-occupied modes and $ |\varphi_q(0)|^2 \ll 1 $ for the highly occupied  
modes, see eq. (\ref{OmegacaseI}).] For the highly occupied modes we 
have $ |\dot\varphi_q(0)|^2 = O(1/g) $. Thus, for early times $ \tau =
{\cal O}(1) $ these modes will have $ |\varphi_q(\tau)|^2 = O(1/g)
$. This makes the correlation function be of order $ 1/g $ near the
origin for early times, as we see in Fig. \ref{frcorrI2}.

\subsubsection{Case II}

In this case at $ \tau = 0 $, the modes have $ |\varphi_q(0)|^2 = O(1/g) $.
This makes the correlation function be of order $ 1/g $ near the
origin for $ \tau = 0 $, as we see in Fig. \ref{frcorrII0} and 
\ref{frcorrIIeta0}.

\subsection{Late time}

The late time behavior of the correlation function depends whether the symmetry
is dynamically broken or not. However, its behaviour is the same for
both cases, I and II. 

We have to distinguish the two regimes:

\subsubsection{ $ {\cal M}_\infty^2 > 0 $ }

In this regime we have the same results for $ \zeta_0 = 0 $ and for 
$ \zeta_0 \ll 1 $.

\begin{figure}[h] 
\epsfig{file=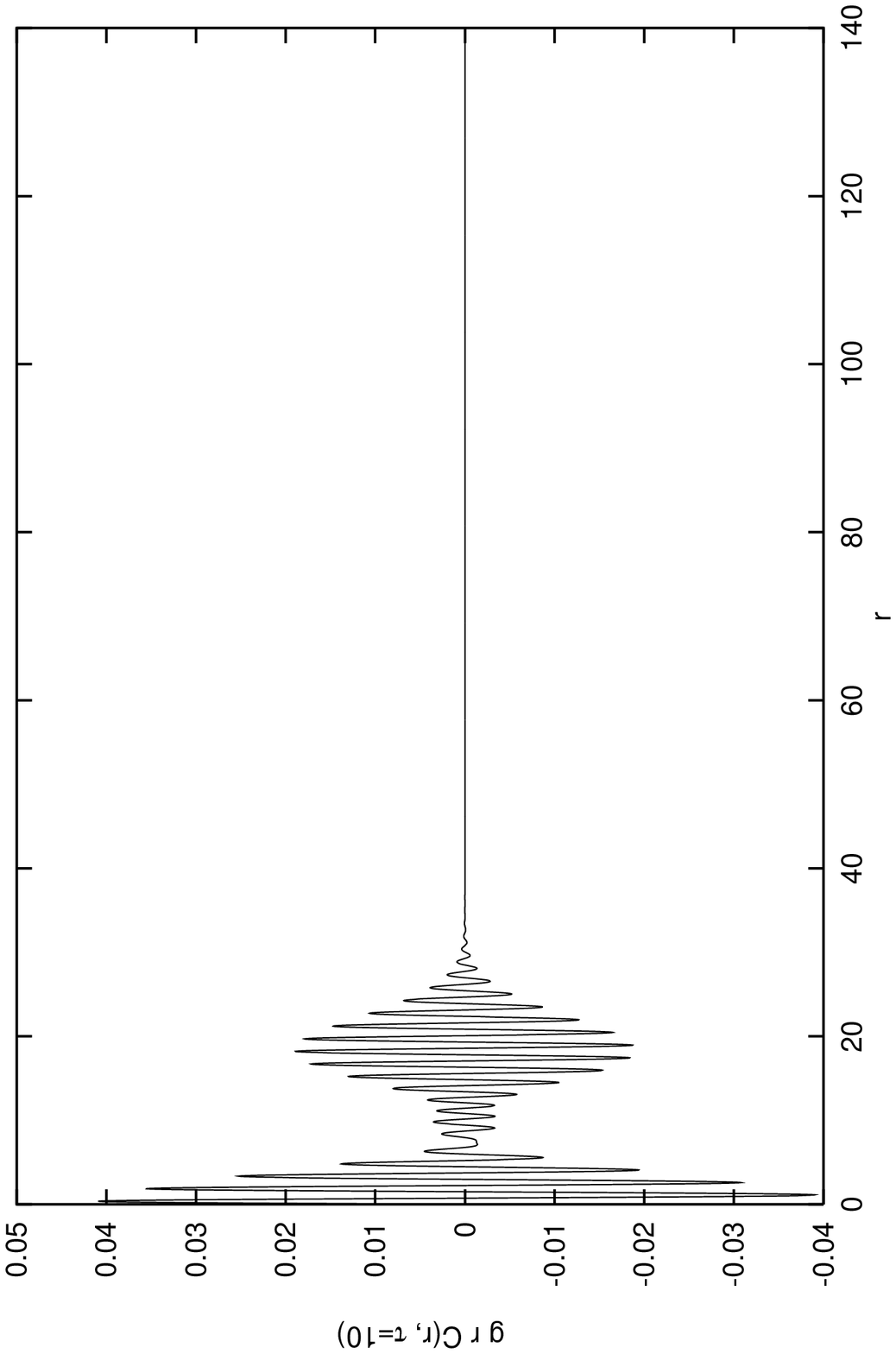}
\caption{Case I. $ m_R^2 > 0 $. Unbroken symmetry.
$ g r C(r, \tau=10) $ with the same $ g $ and initial conditions as in 
Fig. \ref{frcorrI2} .} 
\label{frcorrI10}
\end{figure}

\begin{figure}[h] 
\epsfig{file=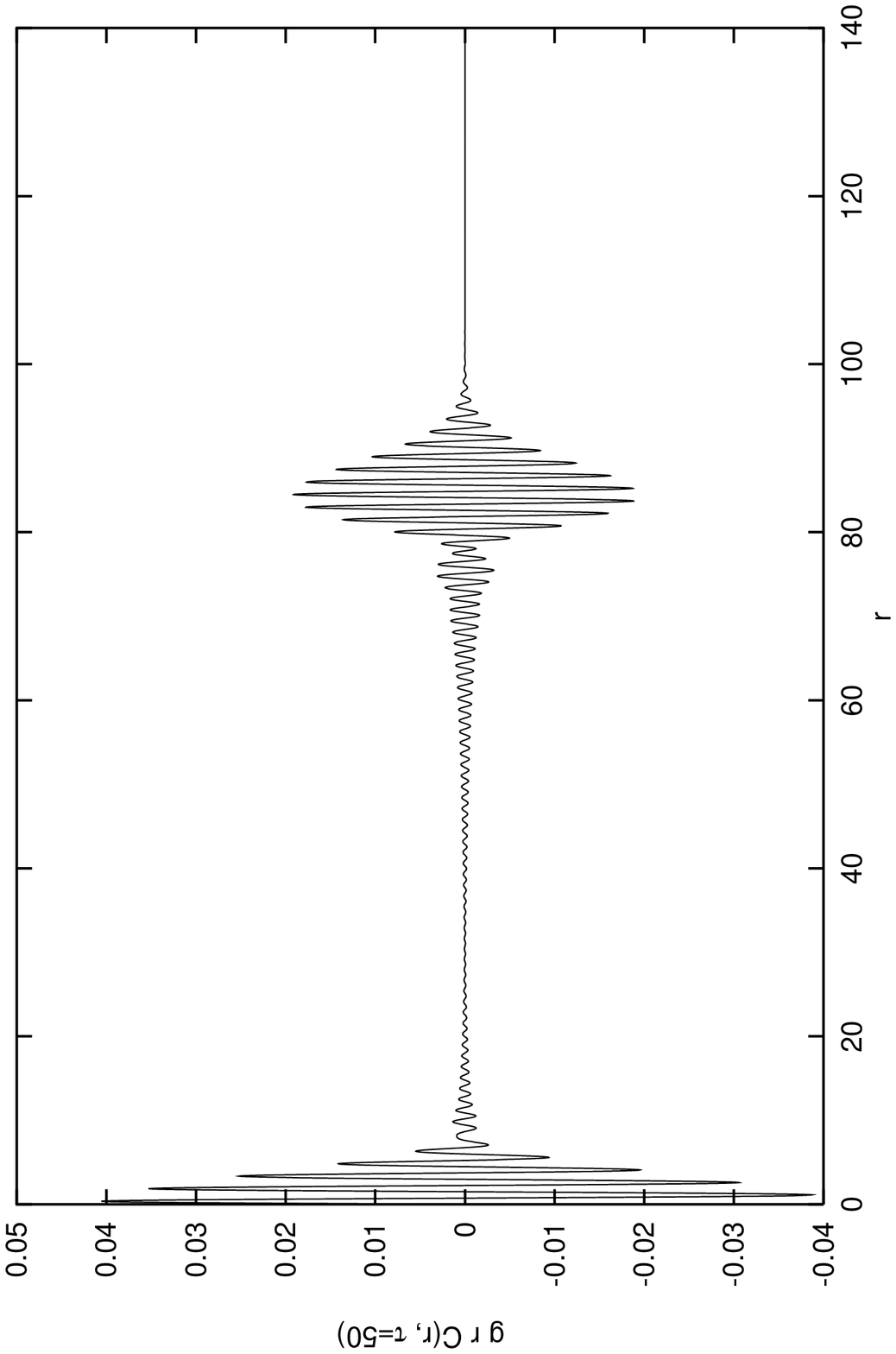}
\caption{Case I. $ m_R^2 > 0 $. Unbroken symmetry. 
$ g r C(r, \tau=50) $ with the same $ g $ and initial conditions as in 
Fig. \ref{frcorrI2} .} 
\label{frcorrI50}
\end{figure}

\begin{figure}[h] 
\epsfig{file=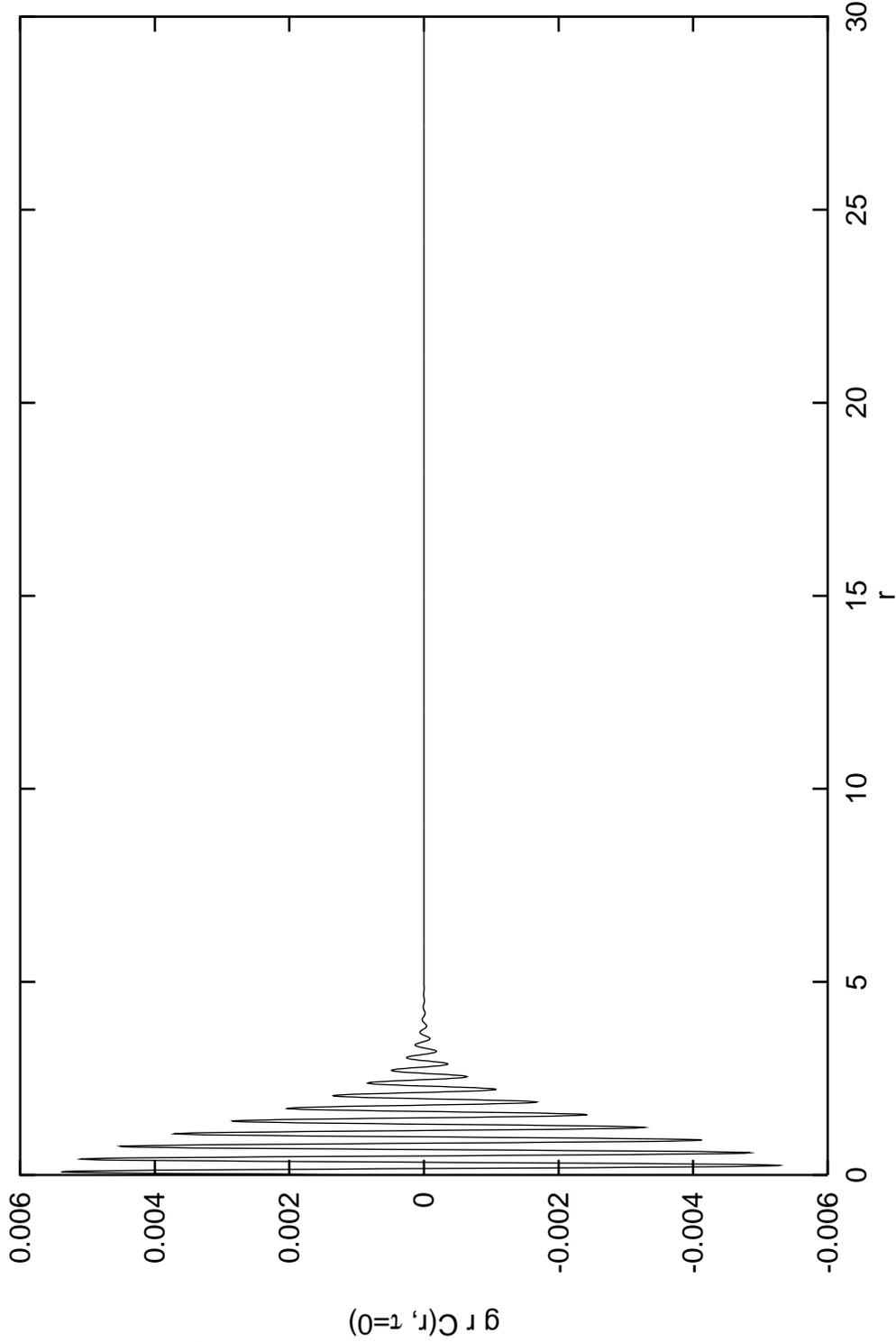}
\caption{Case II. $ m_R^2 < 0 $. Dynamically broken symmetry.
$ g r C(r, \tau=0) $ for $ g = 10^{-7} $ and initial conditions: 
$ \zeta_0 = 0 $, $ gN_0 = 4.478 $, $ q_0 = 1.3083 $, $ \sigma = 0.07850 $.} 
\label{frcorrII0}
\end{figure}

We observe that at intermediate times a spherical pulse develops.
This spherical pulse propagates with a constant radial speed 
given by the radial group velocity for $ q = q_0 $ and an amplitude
that decreases as $ 1/r $. The radial width of the pulse, $ L $,  remains
approximately constant (see Figs. \ref{frcorrI10} and \ref{frcorrI50}).

The group velocity is asymptotically given by,
\begin{equation}\label{vgrupo}
v_g = \left. \frac{d \omega_{q\infty}}{dq} \right|_{q=q_0} =
\left. \frac{d\;}{dq} \sqrt{q^2+{\cal M}_{\infty}^2}\; \right|_{q=q_0} =
\frac{q_0}{\sqrt{q_0^2+{\cal M}_\infty^2}} \; .
\end{equation}

We recall that here $ {\cal M}_\infty^2 = \alpha + \frac{g\Sigma_{max}}{2}
> 1 $ [see eqs. (\ref{Minfmpos}) and (\ref{Minfmnegunbr})]. 
[$ \alpha = {\rm sign}(m_R^2) = \pm 1 $.]

Asymptotically, the correlation function becomes the sum of two terms
\begin{equation}
C(r,\tau) = C_{origin}(r) + C_{p}(r,\tau) \; ,
\end{equation}
where $ C_{origin}(r) $ is the correlation function near the
origin. This term is  asymptotically  time-independent.

The pulse contribution to the correlation function, $ C_{p}(r,\tau) $,
has approximately the asymptotic form, 
\begin{equation}
C_p(r,\tau) = \frac1{g \,r} \; P(r- 2\,v_g\,\tau-c) \; ,
\end{equation}
here $ c $ is a constant of order one, and $ P(u) $ is of the order $
{\cal O}(1)   $ only for $ -L/2 < u < L/2 $ where $ L $ is the width
of the pulse ({\emph i.e.} the pulse is localized around $ r \simeq
2\,v_g\,\tau + c $). 

The pulse term is due to the particles in the initial distribution
that effectively propagate as free particles. This is so since the
effective mass in the mode equations (\ref{kmodes}) becomes
asymptotically constant and hence the modes effectively decouple from
each other. The pulse term is absent when there are no particles in the
initial state \cite{noscorre}.

In summary, the correlator is of order $ {\cal O}(1/g) \gg 1 $ for 
$ 2\,v_g\,\tau + c - L/2 < r < 2\,v_g\,\tau + c +  L/2 $
whereas causality makes it to fall to $ {\cal O}(1) $ values for $ r >
2\,v_g\,\tau + c + L/2 $.   

\subsubsection{ $ {\cal M}_\infty^2 = 0 $ }

In this regime the symmetry is broken and there were spinodal
resonances for earlier times.

Since the effective mass vanishes asymptotically the mode with $ q = 0 $ 
behaves as
\begin{equation}
\varphi_0(\tau) =^{\tau \to \infty} L + K \tau
\end{equation}
The Wronskian guarantees that neither of the complex coefficients $ L $, 
$ K $ can vanish \cite{noscorre}.
This linear growth with time can be interpreted as an out of equilibrium
novel form of Bose Einstein condensation.
We analyse below the contribution of this condensate $ C_s(r,\tau) $ to
the correlation function.

\medskip

We have studied both $ \zeta_0 = 0 $ and $ \zeta_0 \neq 0 $ 
(with $ \zeta_0 \ll 1 $).

\medskip

\paragraph{ $ \zeta_0 = 0 $. }
For late time ($ \tau \gg \tau_s $), the particles created due to
the spinodal resonance contribute to the correlation function with a
term $ C_s(r,\tau) $ of order $ 1/g $  when $ r $ is in the interval $
(0, \, 2\tau ) $. This term decays as $ 1/r $.

\begin{figure}[h] 
\epsfig{file=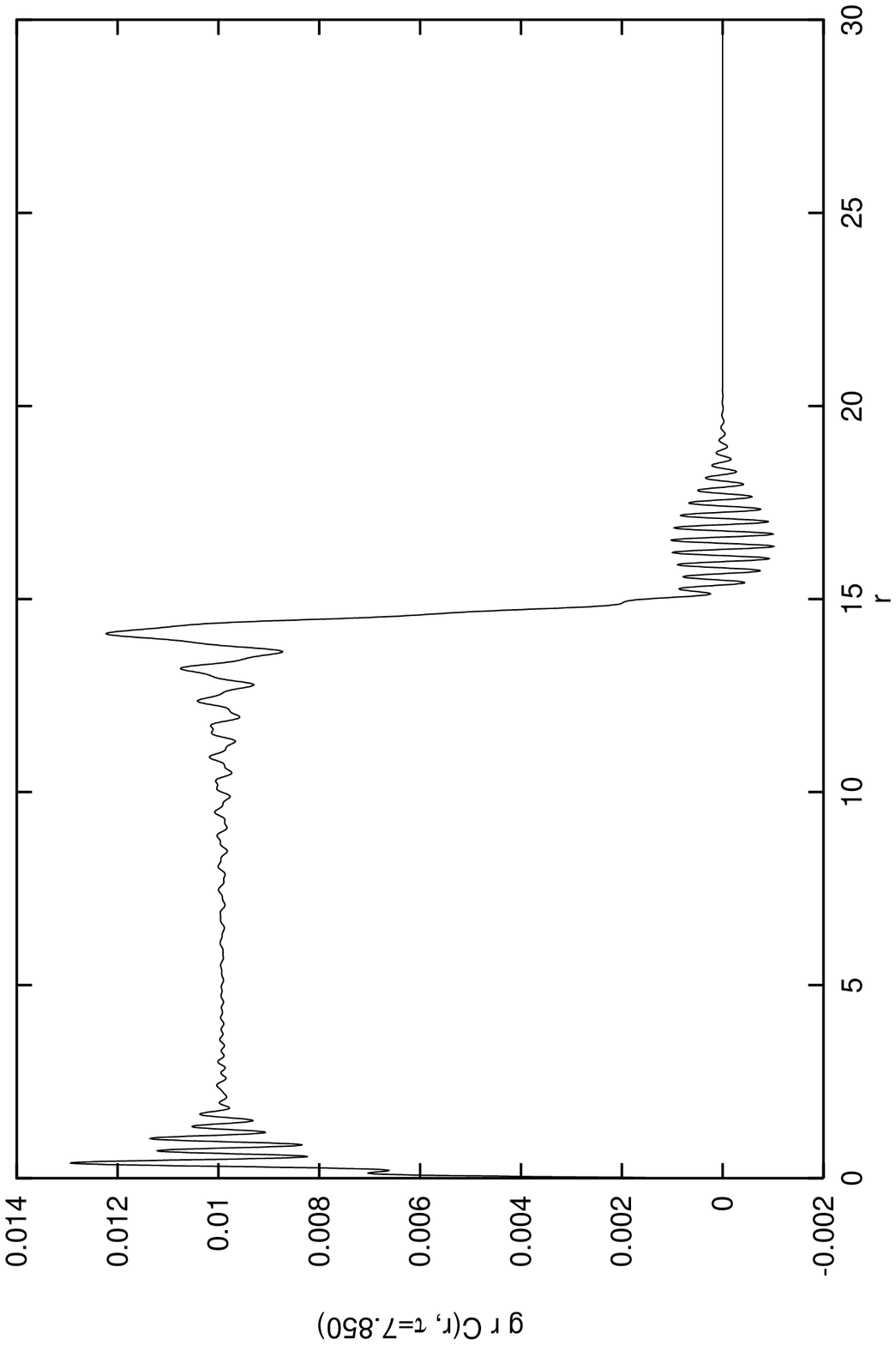}
\caption{Case II. $ m_R^2 < 0 $. Dynamically broken symmetry.
$ g r C(r, \tau=7.850) $ with the same $ g $ and initial conditions as in 
Fig. \ref{frcorrII0} .} 
\label{frcorrII30}
\end{figure}

\begin{figure}[h] 
\epsfig{file=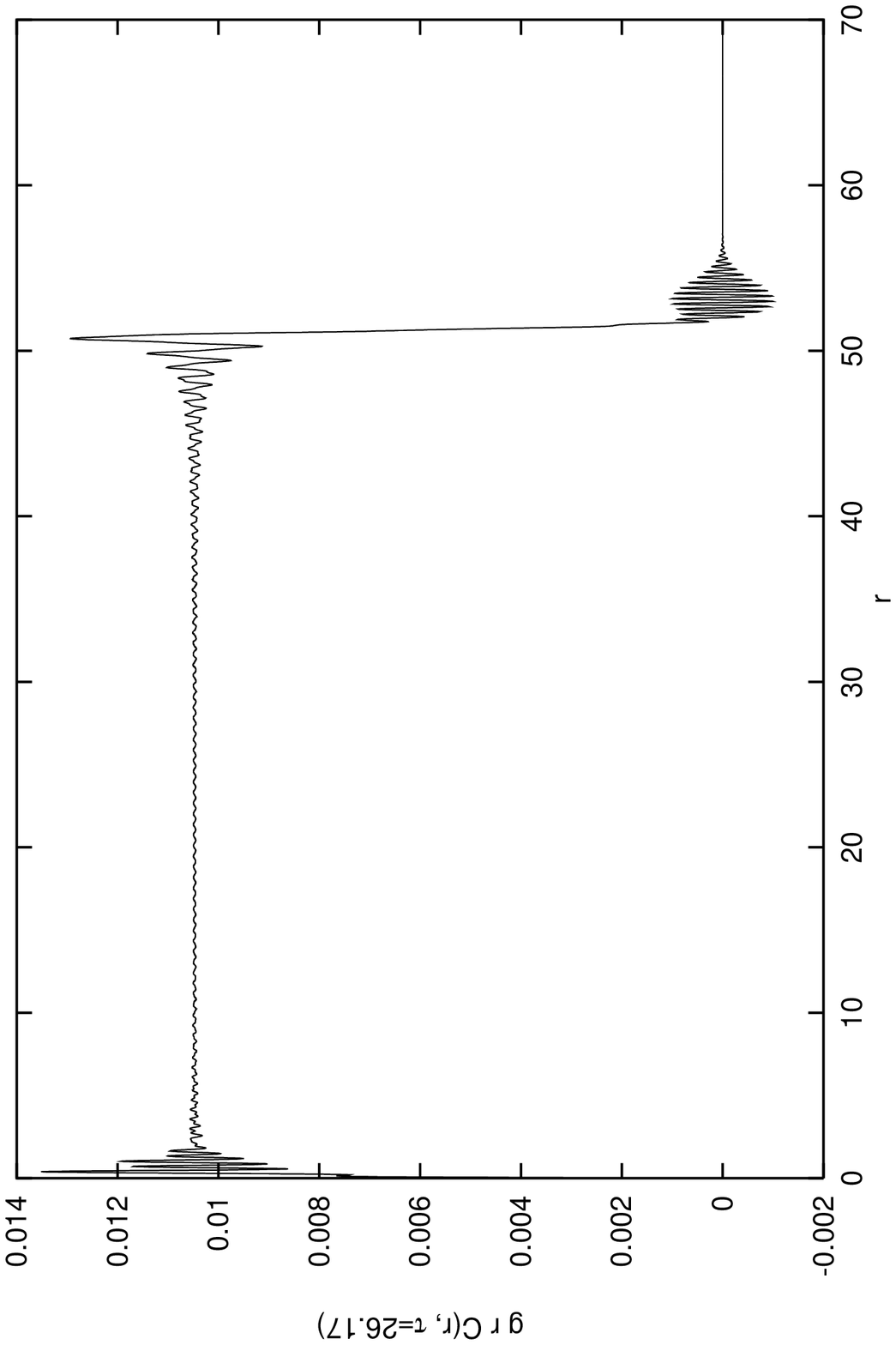}
\caption{Case II. $ m_R^2 < 0 $. Dynamically broken symmetry.
$ g r C(r, \tau=26.17) $ with the same $ g $ and initial conditions as in 
Fig. \ref{frcorrII0} .} 
\label{frcorrII100}
\end{figure}

\begin{figure}[h] 
\epsfig{file=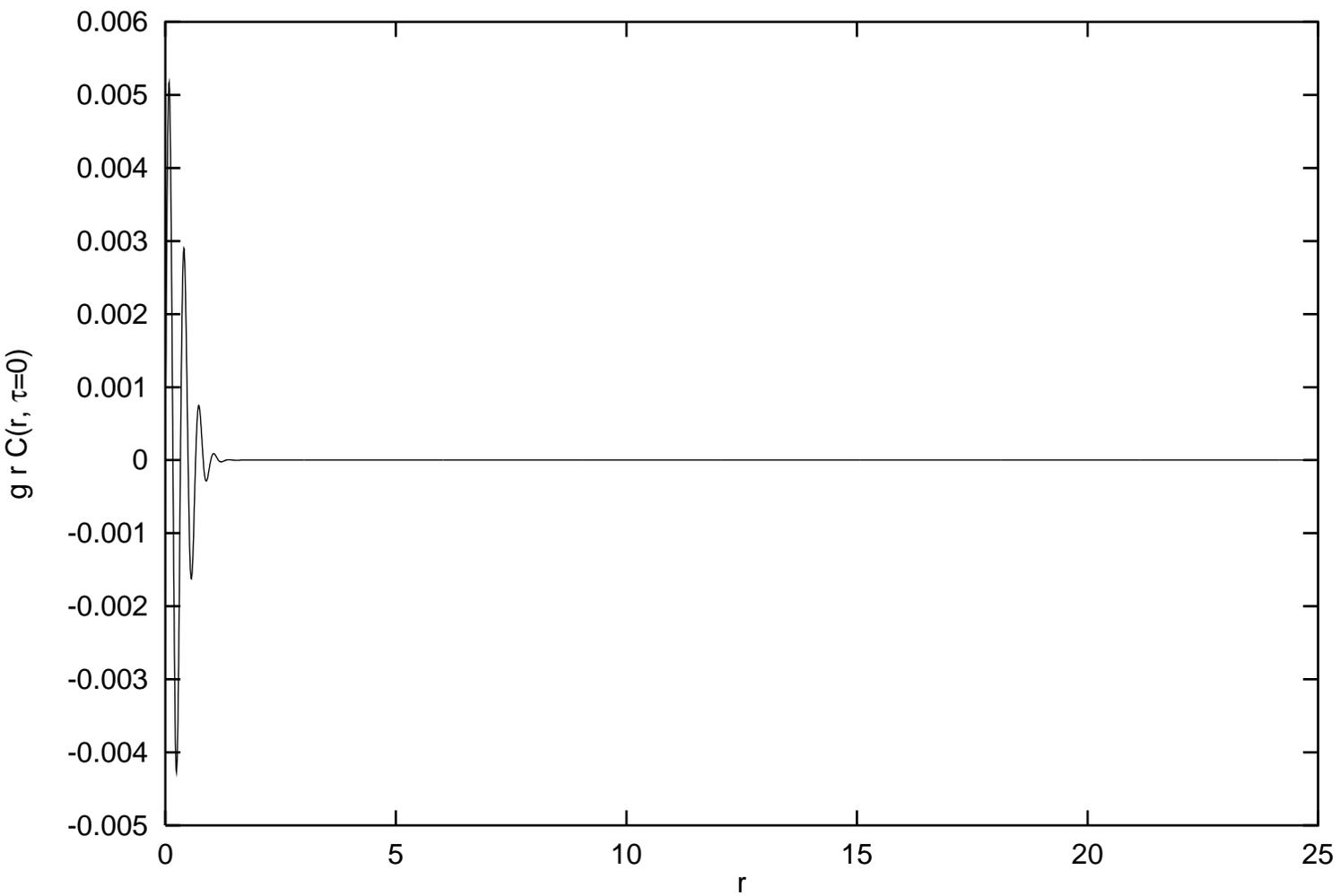}
\caption{Case II. $ m_R^2 < 0 $. Dynamically broken symmetry.
$ g r C(r, \tau=0) $ for $ g = 10^{-7} $ and initial conditions: 
$ \zeta_0 = 0.2617 \cdot 10^{-5} $,
$ gN_0 = 4.478 $, $ q_0 = 1.3083 $, $ \sigma = 0.2617 $.} 
\label{frcorrIIeta0}
\end{figure}

There is in addition a pulse term, $ C_p(r,\tau) $, in the correlator
that moves away from the origin with unit velocity (remember
eq. (\ref{vgrupo}) and that $ {\cal M}_\infty^2 = 0 $). See
Figs. \ref{frcorrII0}-\ref{frcorrII100}. 

Thus, the correlation function is asymptotically given by
\begin{equation}
C(r,\tau) = C_{origin}(r) + C_s(r,\tau) + C_p(r,\tau)
\end{equation}
Here, $ C_{origin}(r) $, the correlation near the origin is 
asymptotically time-independent.

The contribution of the pulse, $ C_p(r,\tau) $,  has the form,
\begin{equation}
C_{p}(r,\tau) = \frac1{g \,r} \; P(r- 2\,\tau-c) \; ,
\end{equation}
The contribution from the particles 
created by spinodal resonance, $ C_s(r,\tau) $, has the form:
\begin{equation}
C_s(r,\tau) = \frac{K}{g \, r} \; Q\!\left( \frac{r}{2\tau} \right) \; .
\end{equation}
where $ K $ is a constant and $ Q(u) = \theta(1-u) $.

Introducing the dynamical correlation length $ \xi(\tau) \sim 2\tau $
and defining the variable $ u = \frac{r}{\xi(\tau)} \; $, $  C_s(r,\tau) $
can be written in the form,
\begin{equation}
C_s(r,\tau) = \frac{K}{g \, u\; \xi(\tau)}\;  \theta(1-u) \; .
\end{equation}
Using the customary notation for the scaling regime 
\begin{equation}
C_s(r,\tau) = \frac1{[\xi(\tau)]^{2(1-z)}} \; I(u) \; ,
\end{equation}
with the anomalous dynamical exponent $ z = 1/2 $ (the naive scaling length
dimension of the field is $ 1 $). In this case the scaling function
is  given by, 
\begin{equation} \label{Itheta}
I(u) = {K \over g \; u } \; \theta (1-u) \; .
\end{equation}

An analogous spinodal term $ C_s(r,\tau) $ in the correlator have been
obtained for initially broken symmetry and no particles in the initial
state in \cite{noscorre}. 

The spinodal term $ C_s(r,\tau) $ can be interpreted  as  follows. For
times $ \tau $ later than   $ \tau_s $ there 
is a zero momentum condensate formed by Goldstone bosons travelling at
the speed of light and  back-to-back. That is, massless particles
emitted from the points   $  (0,\tau) $ and $ (r,\tau) $
form  propagating fronts which at time $ \tau $ are at a distance $
\tau -\tau_s $ from the origin and from  $ r $, respectively.  
These space-time points are causally connected for
$  2(\tau-\tau_s) \geq r $. Otherwise,  $ C_s(r,\tau) $ is not of order
$ 1/g $ but of order one. 

An alternative interpretation of the causality step function goes as
follows. Signals are emitted at the speed of light from all points in
the condensate. We have causal connection between the points $ 0 $ and
$ r $ once signals starting from a given point arrive to both
points. The earlier this happens is for the signals emitted from the
half-away point at $ r/2 $. These signals need a time $ \tau = r/2 $
to reach both points. Hence, the correlator is of the order $ 1/g $
for $ 2 \, \tau > r $. 

Moreover, the analytical derivation of eq. (\ref{Itheta}) from the
low-$q$ behavior of the mode functions given in ref.\cite{noscorre}
also applies here. Notice that the scaling contribution of the
Goldstone bosons to the correlator is {\bf different} from a free
massless propagator. We have here a $ 1/r $ falloff whereas a free
massless scalar field has a $ 1/r^2 $ falloff.

\bigskip

\begin{figure}[h] 
\epsfig{file=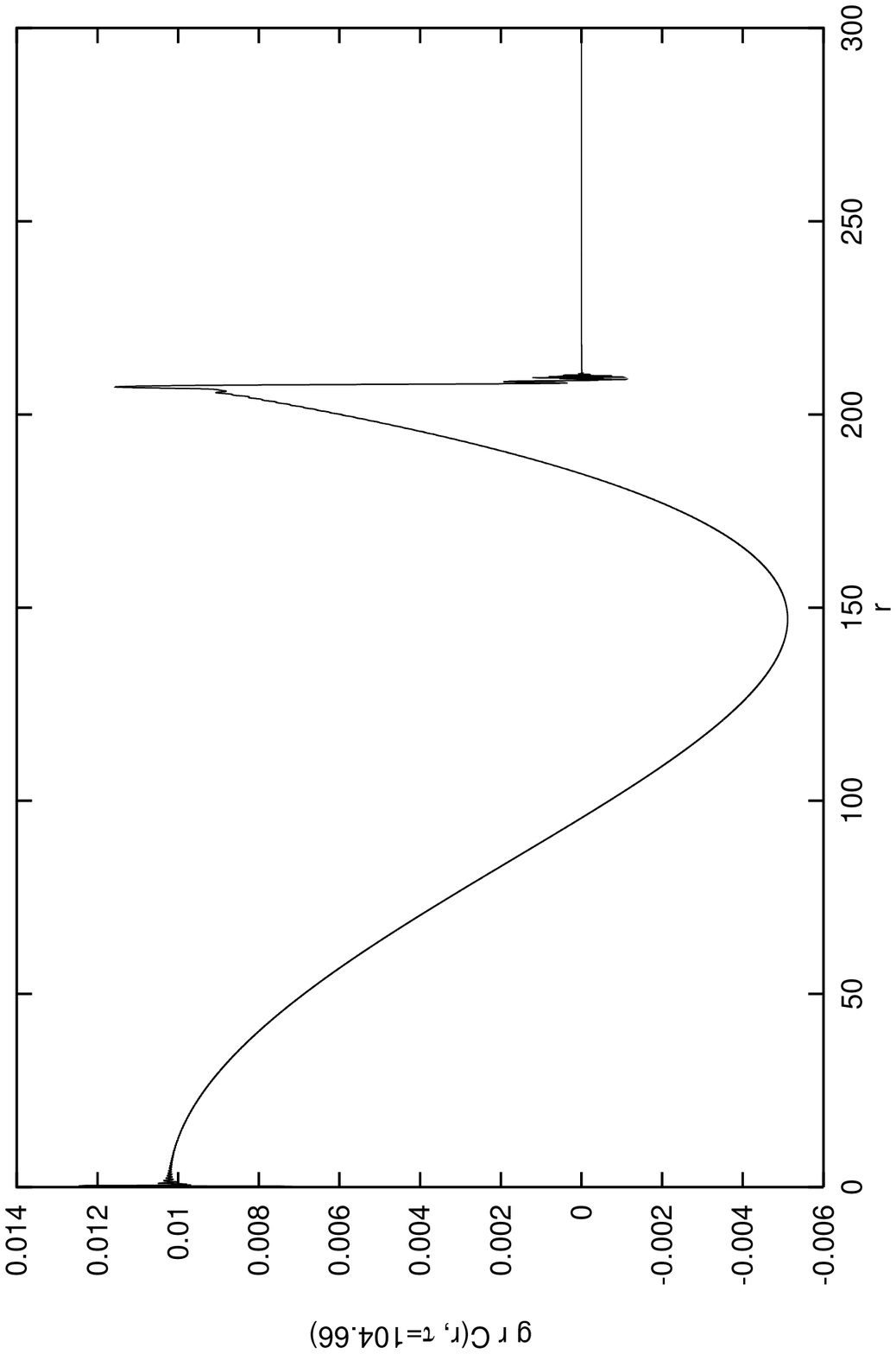}
\caption{Case II. $ m_R^2 < 0 $. Dynamically broken symmetry.
$ g r C(r, \tau=104.66) $ with the same $ g $ and initial conditions as in 
Fig. \ref{frcorrIIeta0} .} 
\label{frcorrIIeta400}
\end{figure}

\begin{figure}[h] 
\epsfig{file=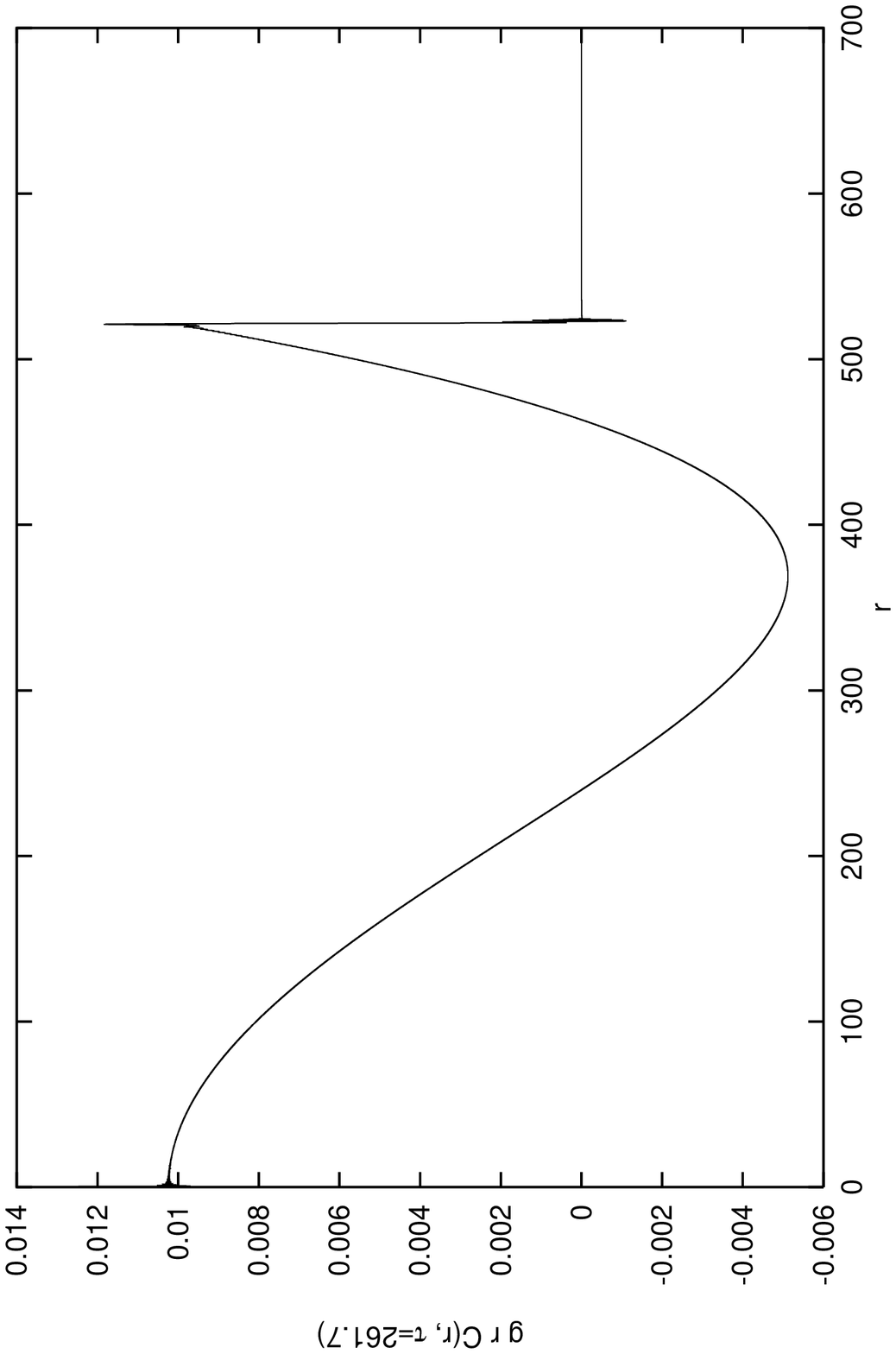}
\caption{Case II. $ m_R^2 < 0 $. Dynamically broken symmetry.
$ g r C(r, \tau=261.7) $ with the same $ g $ and initial conditions as in 
Fig. \ref{frcorrIIeta0} .} 
\label{frcorrIIeta1000}
\end{figure}

\begin{figure}[h] 
\epsfig{file=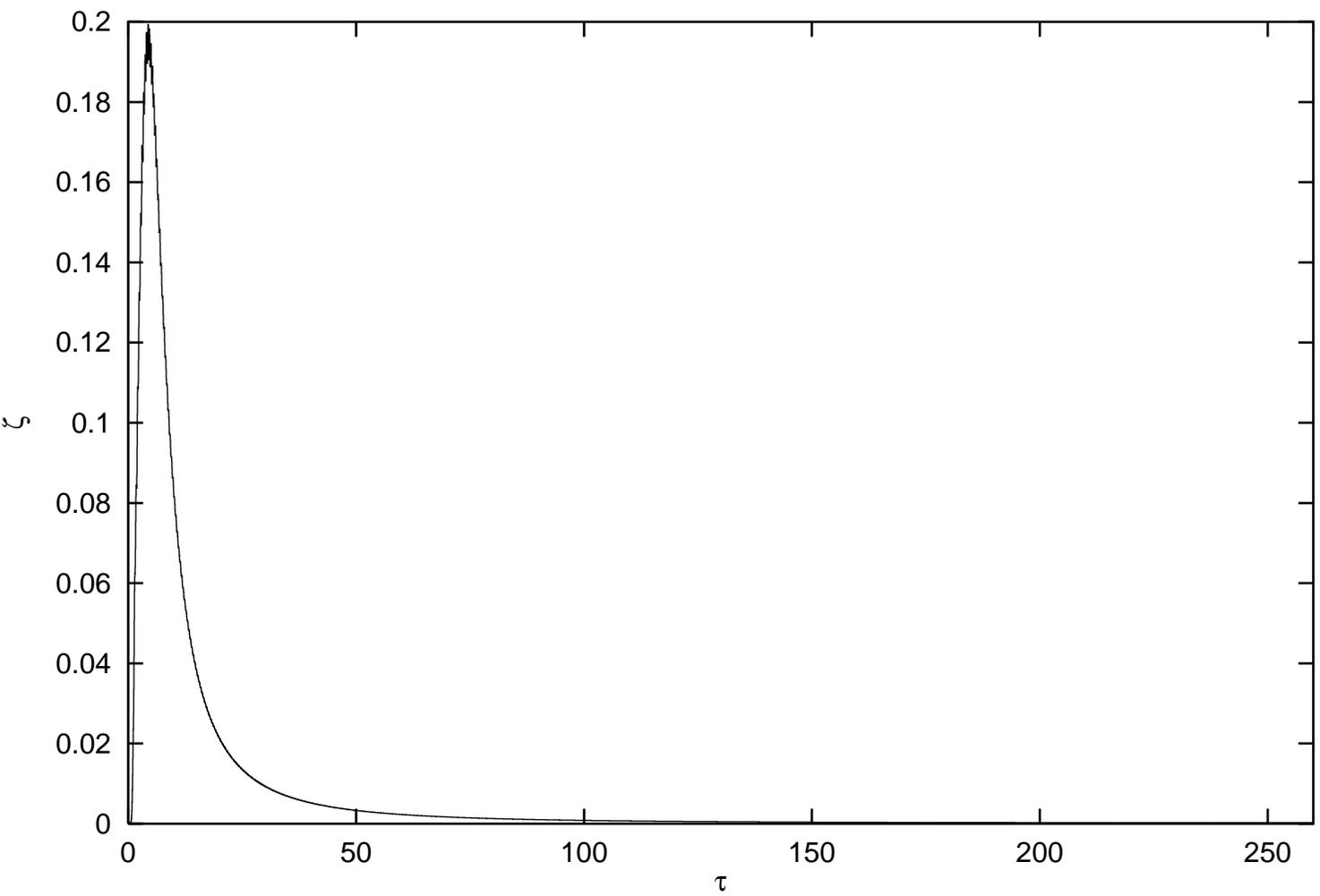}
\caption{Case II. $ m_R^2 < 0 $. Dynamically broken symmetry.
$ \zeta(\tau) $ with the same $ g $ and initial conditions as in 
Fig. \ref{frcorrIIeta0} .} 
\label{fIIeta}
\end{figure}

\paragraph{ $ \zeta_0 \neq 0 $ (with $ \zeta_0 \ll 1 $). }
For late time ($ \tau \gg \tau_s $), the particles created by
spinodal resonance give a contribution $ C_s(r,\tau) $ of order $ 1/g
$ to the correlation function in the interval $ 0 < r < 2\tau $. 
See Figs. \ref{frcorrIIeta0}-\ref{fIIeta}.

The correlation function is asymptotically given by
\begin{equation}
C(r,\tau) = C_{origin}(r) + C_s(r,\tau) + C_p(r,\tau) \; ,
\end{equation}
where the time-independent piece $ C_{origin}(r) $ is the correlation
near the origin.
$ C_p(r,\tau) $ is the contribution of the pulse,
\begin{equation}
C_p(r,\tau) = \frac1{g \,r}\; P(r- 2\,\tau-c) \; ,
\end{equation}
and $ C_s(r,\tau) $ is the contribution from the particles created by
spinodal resonance, 
\begin{equation}
C_s(r,\tau) = \frac1{[\xi(\tau)]^{2(1-z)}} \; I_{\tau}(u) \; ,
\end{equation}
with $ \xi(\tau) \sim 2\tau $ the correlation length, 
$ u = \frac{r}{\xi(\tau)} $, and $ z = 1/2 $.

$ I_{\tau}(u) $ is no longer given by eq. (\ref{Itheta}) when $ \zeta_0 \neq 0
$. $ Q_{\tau}(u) = u\, I_{\tau}(u) $ now oscillates with $ u $. At a
given time $ \tau $, the number of oscillations of $ Q_{\tau}(u) $ in
the interval  $ 0 < u < 1 $ 
equals the number of oscillations performed by the order parameter 
$ \zeta(\tau) $ from  time $ \tau = 0 $ till time $ \tau $. That is,
the scaling exists for $ \zeta_0 \neq 0 $ in a generalized sense since the 
function $ Q_{\tau}(u) $ changes each time  $ \zeta(\tau) $ performs an
oscillation. This is due to the appearance of the extra length scale $
\zeta_0 $. See Figs. 13-16.

\bigskip

As for unbroken symmetry, the pulse term is due to the particles in
the initial distribution that effectively propagate as free particles.

Both for $ \zeta_0 = 0 $ and $ \zeta_0 \neq 0 $ causality makes the
correlator of order $ {\cal O}(1) $ for $ r > 2\,\tau +c $. For $ r <
2\,\tau +c $ the correlations are of order 
$ O(1/g) \gg 1 $.

\begin{appendix}
\section{Adiabatic Approximation for the modes} \label{adiabatic}

In this Appendix we use the adiabatic form of the modes to evaluate
the quantum fluctuations  $ \Sigma(\tau) $.

The modes $ \varphi_q(\tau) $ can be represented as\cite{losala87}
\begin{equation}\label{adiab}
\varphi_q(\tau) = {1 \over \sqrt{{\cal P}_q(\tau)}} \left[ a_q \; e^{-i
\int_0^{\tau} dx \; {\cal P}_q(x) } + b_q \; e^{i \int_0^{\tau} dx \;
{\cal P}_q(x) } \right]
\end{equation}
where $ a_q $ and $ b_q $ are constants and $ {\cal P}_q(\tau) $ depends
on time. Inserting eq. (\ref{adiab}) into eq. (\ref{modeeq}) yields the
following non-linear differential equation for $ {\cal P}_q(\tau) $
\begin{equation}
{{\ddot {\cal P}}_q(\tau)\over 2 \, {\cal P}_q(\tau)} -\frac{3}{4} \left( 
{{\dot {\cal P}}_q(\tau)\over  {\cal P}_q(\tau)}\right)^2 + 
{\cal P}_q^2(\tau) = q^2 + {\cal M}^2(\tau)
\end{equation}
The initial conditions (\ref{inicondq}) combined with the Wronskian
conservation yields 
\begin{equation}\label{condab}
|a_q|^2 -  |b_q|^2 = 1 \; .
\end{equation}
As long as $ {\cal P}_q(\tau) $ is a {\bf real} function, we have for
the squared modulus,
\begin{equation}\label{ficuad}
\left| \varphi_q(\tau) \right|^2 = {1 \over {\cal P}_q(\tau)} \left[ 
|a_q|^2 +  |b_q|^2 + 2  |a_q \; b_q | \cos\left( 2 \int_0^{\tau} dx \;
{\cal P}_q(x) + \alpha_q \right) \right]
\end{equation}
where $  \alpha_q $ is a time independent phase. 

When $ q $ belongs to a parametric resonant band  $ {\cal P}_q(\tau) $
gets an imaginary part. We consider here the case where such
unstabilities have a negligible effect.

We are interested in the modes with large amplitudes $ |a_q| \gg 1 \;
, |b_q| \gg 1 $. That is, those in an interval of width $ \sigma $
around the peak momentum $ q_0 $.  For such modes, thanks to
eq. (\ref{condab}) we can approximate $ |a_q| \simeq |b_q| $. 

Inserting eq. (\ref{ficuad}) into the integral (\ref{sigmaR}) for $ \Sigma(\tau)
$, and approximating the slowly varying factor $ \frac{1}{{\cal P}_q(\tau)} $
by its average on a period (that we will denote 
$ \frac{1}{\widehat{\cal P}_q} $ ) yields 
\begin{equation}\label{sigmadi}
\Sigma(\tau) \simeq 2 \int { q^2 \; dq \over \widehat{\cal P}_q}\;  |a_q|^2
\left[ 1 + \cos\left( 2 \int_0^{\tau} dx \; {\cal P}_q(x) + \alpha_q
\right) \right] 
\end{equation}
It follows from here the bounds
$$
0 \leq  \Sigma(\tau) \leq 4 \int { q^2 \; dq \over \widehat{\cal P}_q}\;
|a_q|^2\equiv \Sigma_{max}
$$

For late times the integral of the oscillating cosinus in
eq. (\ref{sigmadi}) vanishes. Therefore,
$$
 \Sigma(\infty) = 2 \int { q^2 \; dq \over \widehat{\cal P}_q}\; |a_q|^2
$$
and
\begin{equation}\label{desigu}
\Sigma(\infty) = \frac12 \, \Sigma_{max} \; .
\end{equation}

\section{Slowly varying parameters in the late time mode functions}

Let us define the following slowly varying parameters \cite{nos3}:
\begin{eqnarray}
A_q(\tau) &\equiv& \frac{1}{2} \; e^{-i\omega_{q\infty}\tau}
\left[\varphi_q(\tau) - \frac{i}{\omega_{q\infty}} \;
\dot\varphi_q(\tau)\right] \; , \\ 
B_q(\tau) &\equiv& \frac{1}{2} \; e^{+i\omega_{q\infty}\tau}
\left[\varphi_q(\tau)+ \frac{i}{\omega_{q\infty}}
\;\dot\varphi_q(\tau)\right] \; , 
\end{eqnarray}

with,
\begin{equation}
\omega_{q\infty} \equiv \sqrt{q^2+{\cal M}_\infty^2} \quad ; \quad
{\cal M}_\infty^2 \equiv {\cal M}^2(\infty) \; \; .
\end{equation}

These slowly varying parameters are asymptotically constant.

We can express the mode functions in terms of them as follows:
\begin{equation} \label{phiasymp}
\varphi_q(\tau) = A_q(\tau)\; e^{i\omega_{q\infty}\tau} + 
B_q(\tau)\; e^{-i\omega_{q\infty}\tau} \;\; .
\end{equation}

Thus, the squared modulus of the modes is,
\begin{equation}
|\varphi_q(\tau)|^2 = |A_q(\tau)|^2 + |B_q(\tau)|^2 + 
2|A_q(\tau)B_q(\tau)|\cos[2\omega_{q\infty}\tau+\phi_q(\tau)] \; ,
\end{equation}
where
\begin{equation} \label{phasedef}
A_q(\tau)B_q(\tau)^* = |A_q(\tau)B_q(\tau)| e^{i\phi_q(\tau)} \; .
\end{equation}

The constancy of the wronskian implies:
\begin{equation}
|B_q(\tau)|^2 - |A_q(\tau)|^2 = 
\frac{1}{\omega_{q\infty}} \; , \label{wronsrel}
\end{equation}
plus terms with derivatives of the slowly varying parameters that vanish
asymptotically.

We define a slowly varying modulus \cite{nos3},
\begin{equation}
M_q(\tau) \equiv \sqrt g \sqrt{|A_q(\tau)|^2+|B_q(\tau)|^2} \; .
\end{equation}

The main contributions to the physical quantities comes from the 
modes with occupation numbers of order $1/g$, for these modes 
$|A_q(\tau)|$ and $|B_q(\tau)|$ are of order $1/\sqrt{g}$. 
Therefore, $M_q(\tau)$ is of order $1$ for these modes.
Moreover, eq. (\ref{wronsrel}) implies that,
\begin{equation} 
|B_q(\tau)|^2=|A_q(\tau)|^2 \left[1+O(g)\right] \; ,
\end{equation}
and we can thus approximate the squared modulus as follows:
\begin{equation}
g|\varphi_q(\tau)|^2 = M_q(\tau)^2 \left\{ 1+\cos[2\omega_{q\infty}\tau+
\phi_q(\tau)]\right\} [1+O(g)] \; . \label{approxsquaredmod}
\end{equation}

\bigskip

We obtain from (\ref{phiasymp}) a similar formula for the squared modulus 
of the derivative of the mode function,
\begin{eqnarray}
\dot\varphi_q(\tau) = &&i\omega_{q\infty} \left[
A_q(\tau)\; e^{i\omega_{q\infty}\tau} - 
B_q(\tau)\; e^{-i\omega_{q\infty}\tau} \right] \cr
&&+ \dot{A}_q(\tau)\; e^{i\omega_{q\infty}\tau} + 
\dot{B}_q(\tau)\; e^{-i\omega_{q\infty}\tau} \; .
\end{eqnarray}

Using $ \dot{A}_q(\infty) = \dot{B}_q(\infty) = 0 $ and the same
procedure we have 
used to derive the equation for the squared modulus, we obtain:
\begin{equation}
g |\dot\varphi_q(\tau)|^2 = \omega_{q\infty}^2 \; M_q^2(\infty) \left\{
1 - \cos[2\omega_{q\infty}\,\tau+\phi_q(\tau)] \right\} [1+O(g)] 
[1+O(\tau^{-1})] \; .
\end{equation}

Or equivalently,
\begin{equation}   \label{approxsquaredmoddot}
g|\dot\varphi_q(\tau)|^2 = \omega_{q\infty}^2 \left\{ g|\varphi_q(\tau)|^2 -
2\, M_q^2(\tau)\, \cos[2\omega_{q\infty}\,\tau+\phi_q(\tau)] \right\}
[1+O(g)] [1+O(\tau^{-1})] \; . 
\end{equation}

\section{Calculation of the spinodal time in the presence of the
tsunami} \label{secspinodaltime} 

Here, we obtain the early time solution for the modes in the spinodally 
resonant band, and we estimate the spinodal time, $ \tau_s $, both for 
cases, I and II.

Recall that there is spinodal resonance
provided $ m_R^2 < 0 $ and $ g\Sigma_{max} < 2 $.

\bigskip

Let us assume that the time scale for the development of spinodal
unstabilities, $ \tau_s $, is longer than the time scale of damping of the
oscillations in $ g\Sigma(\tau) $, $ \tau_d $. This happens when the
particle peak is wide enough.

Before entering on the calculation of $ \tau_s $, let us remark that the
numerical calculations show that the main effect of $ g\Sigma(\tau) $ for $
\tau<\tau_s $ (and small $ g $) is giving a constant positive term
$\frac{g\Sigma_{max}}{2}$ to the effective squared mass. In fact, this
turns to be true in general, {\em not only} when the oscillations of
$ g\Sigma(\tau) $ are damped before $\tau_s$. 

We have  for $ \tau_d < \tau<\tau_s $:
\begin{equation}	\label{resomo}
-\mu^2 \equiv M^2_{eff}(\tau) \approx 1+\frac{g\Sigma_{max}}{2} < 0 \; .
\end{equation}

Hence,  an approximate equation for the modes reads,
\begin{equation}
\left(\frac{d^2}{d\tau^2}+q^2-\mu^2\right)\varphi_q(\tau)=0 \; .
\end{equation}

Thus, the modes with $ q $ in the interval between $ 0 $ and $ \mu $ are
spinodally resonant.

We consider initial peaks of particles centered well outside the possible 
resonant bands.
Thus, for a sufficiently narrow initial particle peak there are no
particles in the spinodally resonant band. Therefore the initial
condition for modes in the resonant band are:
\begin{equation}
\varphi_q(0) = \frac{1}{\sqrt \Omega_q} 
= \left( q^2 + |{\cal M}^2(0)| \right)^{-1/4},
\;\;\;\; \dot\varphi_q(0) = -i \sqrt \Omega_q 
= -i \left( q^2 + |{\cal M}^2(0)| \right)^{1/4}.
\end{equation}

The solution of eq. (\ref{resomo}) for these modes is:
\begin{eqnarray}
\varphi_q(\tau) = \frac{1}{2\sqrt{\mu^2-q^2}(q^2+|{\cal M}^2(0)|)^{1/4}}
\left[ \left(\sqrt{\mu^2-q^2}-i\sqrt{q^2+|{\cal M}^2(0)|} \right)
e^{\tau\sqrt{\mu^2-q^2}} \right. \nonumber \\ 
\left. + \left(\sqrt{\mu^2-q^2}+i\sqrt{q^2+|{\cal M}^2(0)|}\right)
e^{-\tau\sqrt{\mu^2-q^2}}\right] \;.  \label{cuadmo}
\end{eqnarray}

We thus obtain for the squared modulus neglecting
the exponentially decreasing term 
\begin{eqnarray}
|\varphi_q(\tau)|^2 &\approx& \frac{1+\mu^2}
{4(\mu^2-q^2)\sqrt{q^2+|{\cal M}^2(0)|}}\; e^{2\tau\sqrt{\mu^2-q^2}}\; \cr\cr
&=& \frac{1+\mu^2}
{4\mu^2\left(1-\frac{q^2}{\mu^2}\right)\sqrt{q^2+|{\cal M}^2(0)|}}\; 
e^{2\tau\mu^2\sqrt{1-\frac{q^2}{\mu^2}}}\; . \label{modphiapprox}
\end{eqnarray}

The contribution of the spinodal band to $ g\Sigma(\tau) $ is given by,
\begin{equation}\label{gsigspi}
\Sigma_s(\tau) = \int_0^\mu \; {q^2\; dq\; |\varphi_q(\tau)|^2} \; . 
\end{equation}
Inserting eq. (\ref{modphiapprox}) into eq. (\ref{gsigspi}) we obtain an
estimation for the spinodal growth of the quantum fluctuations. 

To approximately evaluate this integral, we can make further simplifications
in eq. (\ref{modphiapprox}). As $ q/\mu < 1 $ and the contribution of
the modes with $ q \approx \mu $ is exponentially suppressed,
we can expand in $ q/\mu $ to second order in the exponential and to zeroth
order in the factor outside  the exponential.  
\begin{equation}
|\varphi_q(\tau)|^2 \approx \frac{1+\mu^2}
{4\mu^2\,\sqrt{q^2+|{\cal M}^2(0)|}} \;
e^{2\tau\mu}\;e^{-\tau\frac{q^2}{\mu^2}} \; .
\end{equation}

In addition the integrand has its maximum at $ q = O(0.1\mu) $ and 
$ 0 < \mu < 1 $. Therefore we can do the approximations $ 1+\mu^2 \sim 1 $ and
$ \sqrt{q^2+|{\cal M}^2(0)|} \sim 1 $ (for both case I and case II).
Thus,
\begin{equation}
|\varphi_q(\tau)|^2 \approx \frac{1}{4\mu^2} \;
e^{2\tau\mu}\;e^{-\tau\frac{q^2}{\mu^2}} \; .
\end{equation}
Then the integral over $ q $ takes the value
\begin{equation}
\Sigma_s(\tau) \approx \frac{\sqrt\pi\,\mu}{16}\; 
\frac{e^{2\tau\mu}}{\tau^{3/2}} \; .
\end{equation}

The spinodal time $ \tau_s $ is by definition, the time where the
unstabilities are shut off for all $ 0 \leq q \leq \mu$. This happens
when the spinodal modes contribution $ \Sigma_s(\tau) $ compensates the
initial (negative) value of  $  M^2_{eff}(\tau) $ [see eq. (\ref{resomo})]. 
\begin{equation}
g\Sigma_s(\tau_s) \approx \mu^2
\end{equation}
Therefore  $ \tau_s $ is given by the following implicit equation,
\begin{equation}\label{tauese}
\tau_s = \frac{1}{2\mu}\, \log\left[\frac{16}{g\,\sqrt{\pi\,\mu}}\right] 
+ \frac{3}{4\mu}\log(\mu\tau_s) \;.
\end{equation}

The spinodal times given by this equation are in good agreement with the
numerical results.

\section{The initial quantum fluctuations in case II}

The quantum fluctuations at initial time $ $ are related with the
initial data $ $ through eq.(\ref{gsigmaII0})
$$
[g\Sigma_{II}(0)]^3 + (q_0^2+\alpha)[g\Sigma_{II}(0)]^2 -
\left(\frac{gN_0}{\pi}\right)^2 = 0 \quad .
$$
For $ \frac{gN_0}{2\, \pi} > \left({q_0^2+\alpha \over 3}
\right)^{3/2} $ the positive solution of this equation takes
the form, 
$$
g\Sigma_{II}(0) = -{q_0^2+\alpha \over 3} + s_+ + s_-
$$
where
$$
s_{\pm} = \frac1{2^{1/3}} \left( \frac{gN_0}{\pi}\right)^{2/3}\left[ 1
- \frac12 \left( \frac{2\pi}{gN_0}\right)^2 \left({q_0^2+\alpha \over
3}\right)^3 \pm \sqrt{1 - \left( \frac{2\pi}{gN_0}\right)^2
\left({q_0^2+\alpha \over 3}\right)^3}\right]^{1/3}
$$

For $\frac{gN_0}{2\, \pi} < \left({q_0^2+\alpha \over 3}
\right)^{3/2} $ the positive solution can be written as,
$$
g\Sigma_{II}(0) = {q_0^2+\alpha \over 3} \left( 2 \, \cos \beta
-1\right)
$$
with
$$
\cos 3\beta = 2 \, \left({3\over q_0^2+\alpha}\right)^3 \left
( \frac{gN_0}{2\pi}\right)^2 -1 \quad . 
$$

In limiting cases we recover eq.(\ref{casolim}).

\end{appendix}

\acknowledgements

We thank D. Boyanovsky for useful discussions. F. J. C. thanks the
Ministerio de Educaci\'on y Cultura (Spain) for financial support through 
the program Becas de Formaci\'on de Profesorado Universitario en el Extranjero.

\end{document}